\documentclass[a4paper]{article}
\usepackage[margin=25mm]{geometry}
\usepackage{graphicx}
\usepackage{amsmath} 
\usepackage{amssymb}
\usepackage{makecell}
\usepackage{xcolor}
\usepackage{dcolumn,verbatim}
\usepackage{multirow}  
\usepackage[none]{hyphenat} 
\usepackage{chngcntr} 
\usepackage{float} 
\usepackage[hidelinks]{hyperref} 
\usepackage{arydshln} 

\graphicspath{ {./fig/} }
\counterwithout{equation}{section} 

\title{Finding the Optimal Currency Composition of Foreign Exchange Reserves with a Quantum Computer}
\author{Martin Vesel\'y
\thanks{Contact address: martin.vesely@cnb.cz. This work was supported by Czech National Bank Research Project No. P2/22.
The author would like to thank Simona Malovan\'a and Martin Hodula for invaluable advice and project supervision.
The views expressed in this paper are those of the author and not necessarily those of the Czech
National Bank.}
\\Czech National Bank - Risk Management Department
}

\begin{document}
\interfootnotelinepenalty=10000

\maketitle

\begin{abstract}
Portfolio optimization is an inseparable part of strategic asset allocation at the Czech National Bank. Quantum computing is a new technology offering algorithms for that problem. The capabilities and limitations of quantum computers with regard to portfolio optimization should therefore be investigated. In this paper, we focus on applications of quantum algorithms to dynamic portfolio optimization based on the Markowitz model. In particular, we compare algorithms for universal gate-based quantum computers (the QAOA, the VQE and Grover adaptive search), single-purpose quantum annealers, the classical exact branch and bound solver and classical heuristic algorithms (simulated annealing and genetic optimization). To run the quantum algorithms we use the IBM Quantum\textsuperscript{TM} gate-based quantum computer. We also employ the quantum annealer offered by D-Wave. We demonstrate portfolio optimization on finding the optimal currency composition of the CNB's FX reserves. A secondary goal of the paper is to provide staff of central banks and other financial market regulators with literature on quantum optimization algorithms, because financial firms are active in finding possible applications of quantum computing.
\end{abstract}

\section{Introduction and Motivation}
\label{secIntro}
Over the last few years, knowledge of quantum computing has been penetrating the finance industry \cite{qubo-fin-arbitrage,finance-general-overview}. This development has naturally attracted the attention of central banks and regulators \cite{finance-general-riksbank}. As \cite{finance-general-riksbank} is rather theoretical, we tried to assess the practical capabilities of quantum computers ourselves \cite{finance-general-my}. Although the results concerning portfolio optimization with a quantum linear equation solver (the HHL algorithm) and risk evaluation with quantum Monte Carlo  were discouraging, a quantum heuristic for quadratic unconstrained binary optimization (QUBO) -- employed again in portfolio optimization -- worked as expected. We therefore decided to further investigate quantum QUBO techniques. We will demonstrate them and assess their capabilities using dynamic portfolio optimization, in particular optimization of the currency composition of the CNB's FX reserves, a highly important part of its strategic asset allocation process. Because of the latter, the paper contributes to the discussion on optimal reserve currency composition, although our purely technical approach to optimization is not intended as a replacement for expert judgment. On top of that, the article serves as a practical guide to quantum QUBO algorithms for staff of central banks and other financial market regulators.

Before we start our investigation of quantum QUBO algorithms in connection with portfolio optimization, it is worth listing other possible applications of QUBO in finance and general business. There are many problems which can be converted to optimization in a graph and finally to quantum QUBO. \cite{qubo-ising-formulations} provides a (non-exhaustive) list. An outstanding example of graph optimization is the travelling salesperson problem, which can be employed to find arbitrage opportunities, as shown by \cite{qubo-fin-arbitrage}. Besides graph problems, systems of linear equations often arise in real world problems. Since the HHL algorithm designed by \cite{misc-HHL-alg} seems to be far from practical deployment, as noted by \cite{misc-hhl-issues} and \cite{misc-hhl-issues2}, a linear system solver leveraging quantum QUBO was proposed by \cite{qubo-linear-equations}. Linear regression, another technique often used in finance, is a particular application of linear systems. The evaluation of regression coefficients with quantum QUBO is discussed by \cite{qubo-regression}. Linear systems can also be employed in the numerical solution of partial differential equations in finance, for example in option pricing. For this purpose, a modification of a quantum QUBO heuristic was proposed by \cite{qubo-fin-pde}. An example of the application of quantum computers in macroeconomics was presented by \cite{qubo-rbc}, who designed a QUBO-based algorithm for finding the optimal choice between consumption and the capital level in a real business cycle model. Interestingly, quantum QUBO for factoring integers to prime constituents was presented by \cite{qubo-factoring} as a temporary replacement for the algorithm designed by \cite{misc-shor-alg}, which currently suffers from quantum hardware deficiencies. Although integer factorization is not relevant for FX reserves management, it is important for the security departments of banks and other institutions, because quantum factoring algorithms could potentially break the RSA ciphering mechanism. This shortlist of possible applications shows the importance of QUBO in finance and other fields. Interested readers can find a vast overview of real-world QUBO applications outside finance (such as the pharmaceuticals and logistics industries) employing quantum approaches in \cite{qubo-industry-appl}. 

The focal point of this article, however, is portfolio optimization. There are several studies on portfolio optimization employing quantum QUBO. In \cite{ptf-optim-abudhabi}, it is simply stated that portfolio optimization in Markowitz-like fashion can be carried out on a quantum computer. The authors of the study used binary flags only, i.e.~they found out whether to invest in a particular asset or not. \cite{ptf-optim-trajectory} worked with integer asset weights and showed how to impose constraints via a penalty function. \cite{ptf-optim-vol} demonstrated a technique for reducing a higher-order objective function, arising as a result of portfolio volatility targeting, back to a quadratic one solvable with QUBO algorithms. In contrast to previous studies, the authors used real-world data on the S\&P 100 and S\&P 500 equity indices. \cite{ptf-optim-dynamic} added the time dimension to make the portfolio optimization dynamic. Additionally, they provided a performance comparison of some quantum techniques with classical ones and demonstrated the superiority of the quantum approach for problems with more than 20 binary variables. Finally, \cite{ptf-optim-tracking} presented a non-convex QUBO model for replication of a benchmark with a specified number of securities from the benchmark (the ``cardinality constraint'').

To perform our QUBO problem, we will first use methods for the simulation of spin-glasses (the Ising model) on the IBM Quantum\textsuperscript{TM} universal gate-based quantum computer, specifically the Quantum Approximate Optimization Algorithm (QAOA) designed by \cite{qaoa-orig-paper} and the Variational Quantum Eigensolver (VQE) proposed by \cite{vqe-orig-paper}. We will also employ a single-purpose quantum computer intended only for solving QUBO problems – the \textit{quantum annealer} provided by D-Wave. These three techniques are known as the \textit{adiabatic approach}. 

As pointed out by \cite{qubo-preskill}, adiabatic approach-based algorithms seem to be the first real-world applications of quantum computers. However, their main disadvantage is unproven speed-up. Despite this fact, several studies \cite{qubo-fin-crises,qubo-routing,qaoa-warm-start,ptf-optim-dynamic,qubo-rbc} indicated that in particular instances the algorithms perform better than, or at least equally well as, their classical counterparts. On the other hand, \cite{annealing-pitfalls} questioned the claimed advantages of the adiabatic approach. The study also emphasized that often only the simulated annealing developed by \cite{classical-annealing-metropolis} is used as a benchmark in the performance assessment of adiabatic approach-based algorithms. \cite{qubo-machine-learning} took a ``Goldilocks'' stance that further research is needed, especially once quantum computers with a higher number of qubits become available. \cite{annealing-d-wave-res} also pointed out the need for further research and provided empirical evidence of the better performance of the D-Wave quantum annealer in hardware-tailored benchmark problems. The discussion above indicates that there is no consensus on the capabilities of the algorithms. To shed more light on this issue, we will assess the performance of adiabatic approach-based algorithms ourselves. We will expand the set of classical algorithms used in the comparison to include the branch and bound exact solver designed by \cite{classical-branch-bound} and the genetic-based heuristic algorithm proposed by \cite{classical-genetic-first1} and \cite{classical-genetic-first2}. To compare the QUBO approach to portfolio optimization with classical continuous quadratic programming, we will also employ the classical continuous gradient method.

Besides the algorithms described above, we will also test the algorithm proposed by \cite{grover-qubo} based on the quantum database searching developed by \cite{grover-orig-paper}. In contrast to adiabatic approach-based algorithms, the Grover algorithm offers proven quadratic speed-up, as does the derived QUBO algorithm. To the best of our knowledge, there is no study concerning the practical capabilities of the Grover QUBO optimizer. We will therefore address this gap in our paper. 

Interestingly, all the studies on portfolio optimization mentioned above work with equity market data. We will use currency market data instead to find out whether there is any impact on the performance of quantum algorithms, particularly those without rigorously proven speed-up. 

We tested all the algorithms discussed above on the currency composition optimization of the CNB's FX reserves. We found that the best performance from the perspective of both run time and ability to find the global optimum is offered by the \textit{hybrid heuristic} (exploiting a combination of the classical and quantum approaches) provided by D-Wave. Moreover, the hybrid heuristic is mature enough to allow us to test a problem of practical size. In the cases of the VQE and the QAOA, we tested only toy models on real quantum hardware. Grover adaptive search was tested only on a simulator due to the limited number of qubits on the available quantum processors. The exact branch and bound algorithm is able to find the solution to the QUBO version of the problem provided that we have enough memory (more than 8~GB RAM), but the run time is still several hours. Interestingly, we found that the gradient solver implemented in MS~Excel outperforms all of the algorithms tested (both the classical and the quantum ones) once the problem is formulated as a continuous problem. Therefore, we concluded that carrying out portfolio optimization with the QUBO approach is not the best option, at least for the time being. However, as portfolio optimization in binary form seems to be a hard-to-solve problem, it can serve as a testing problem for new algorithms or for the measurement of progress in quantum hardware development.

Our results do not lessen the role of quantum computing in finance. We have to bear in mind that quantum computers are not a fully mature technology yet. We need to wait several years, track the development of technology, build up our ``quantum knowledge'' and then pass a final verdict on the usefulness of quantum computers for finance applications. We should also critically assess which financial problems quantum computers are a suitable option for and whether classical computing may offer better algorithms.

The rest of this paper is organized as follows. The second section provides the basics on the application of QUBO in portfolio optimization, the third section introduces the quantum algorithms employed and briefly discusses classical ones, the fourth section details the currency composition optimization problem and discusses the data used and the results, and finally the fifth section concludes. Note that the basics of quantum computing and the necessary mathematical background and notation are provided in our previous article \cite{finance-general-my}.

\section{Quadratic Binary Portfolio Optimization}
\label{secQuboPtf}
In our research, we employ portfolio optimization based on quadratic programming as introduced by \cite{misc-markowitz}. The model assumes real (continuous) weights on the assets. However, quantum computers are not able to work with real variables at their current stage of development. \cite{grover-fmin-ideas} presented a basic idea of how the quantum database search designed by \cite{grover-orig-paper} could be used for minimization of a general function but on a discrete domain. Hence, this algorithm would need too many qubits to approximate a continuous function sufficiently. Moreover, the algorithm assumes the existence of quantum memory (qRAM), which is currently rather an experimental device -- see the discussion in \cite{finance-general-my} for details. \cite{continuous-complexity} showed that a quantum continuous optimization algorithm based on Grover search would reduce the time $T$ needed to find the global extreme of a function in the classical case to $\sqrt{T}\,\ln(T)$. However, the authors did not provide any technical detail on the construction of the algorithm. \cite{continuous-qaoa} introduced a quantum version of the gradient descent method based on the Quantum Approximate Optimization Algorithm of \cite{qaoa-orig-paper}, but the authors stated that implementation of the algorithm is beyond the capabilities of current quantum computers. This short investigation therefore shows that the original continuous quadratic program used in the Markowitz model has to be modified to one with binary variables (in particular QUBO) in order to be carried out on current quantum computers.

In \cite{finance-general-my} we provided a toy model of portfolio optimization with QUBO and found that such optimization runs successfully on IBM Quantum\textsuperscript{TM} processors. In this part, we will expand the toy model to ``real world'' \textit{dynamic portfolio optimization with discrete time steps}. As discussed by \cite{ptf-optim-dynamic}, there are also static and continuous-time versions of portfolio optimization. However, the discrete-time version is of greater practical value because the static version assumes immutability of portfolio composition, a condition that is hardly fulfilled in practice, and the continuous version works with the unrealistic assumptions that the portfolio changes in each infinitesimal time moment and that assets are perfectly divisible.

\subsubsection*{The QUBO Problem in General}

Before turning to the particular application of QUBO in portfolio optimization, it is worth reminding ourselves of the general form of the QUBO problem. Let us denote $\mathbf{A} \in \mathbb{R}^{n,n}$, $\mathbf{b} \in \mathbb{R}^n$ (a column vector), $\mathbf{x} \in \{0;1\}^n$ (a column vector composed of $n$ binary variables) and  $c \in \mathbb{R}$. With this notation, the QUBO problem is defined as
\begin{equation}
\label{eq_qubo_matrix}
\frac{1}{2}\mathbf{x}^T \mathbf{A} \mathbf{x} + \mathbf{b}^T\mathbf{x} + c \rightarrow \min.
\end{equation}
The problem \eqref{eq_qubo_matrix} can be rewritten in ``sum form''
\begin{equation}
\label{eq_qubo_sum}
\frac{1}{2}\sum_{i=1}^n\sum_{j=1}^n a_{ij}x_i x_j + \sum_{i=1}^n b_i x_i + c \rightarrow \min.
\end{equation}
Naturally, the QUBO problem does not involve any constraint. However, in real-world applications, constraints are often required. To incorporate them, a \textit{penalty function} is added to function \eqref{eq_qubo_sum} and increases its value if any constraint is not fulfilled. Techniques for the construction of the penalty function are discussed by \cite{qubo-industry-appl}.

\subsubsection*{Portfolio Optimization with QUBO}

If matrix $\mathbf{A}$ is replaced with the asset return covariance matrix $\mathbf{C} \in \mathbb{R}^{n,n}$ and multiplied by coefficient $\lambda \ge 0$ expressing risk aversion, vector $\mathbf{b}$ is substituted with average asset returns $\mathbf{r} \in \mathbb{R}^n$ and multiplied by -1 to maximize the returns and $c = 0$, we get the optimization problem $-\mathbf{r}^T\mathbf{x} + \lambda \mathbf{x}^T \mathbf{C} \mathbf{x} \rightarrow \min$, which is nearly the portfolio optimization model proposed by \cite{misc-markowitz}. To have the model complete, we have to add a budget constraint $\mathbf{p}^T\mathbf{x} \le B$, where $\mathbf{p} \in \mathbb{R}^n$ is a vector of the money amounts invested in the assets and $B \in \mathbb{R}$ is the total budget. We can consider that the whole budget is allocated, i.e.~$\mathbf{x}^T\mathbf{p}  - B = 0$, which leads to the  optimization problem
\begin{equation}
\label{eq_qubo_markowitz_bin}
- \mathbf{r}^T\mathbf{x} + \lambda \mathbf{x} ^T\mathbf{C} \mathbf{x}  + F(\mathbf{p}^T \mathbf{x} - B )^2 \rightarrow \min,
\end{equation}
where $F$ is the importance of the budget constraint. The value of $F$ should be sufficiently large so that the objective function value is prohibitively high for infeasible solutions.

Still, this binary form of Markowitz optimization only tells us whether to invest in a particular asset ($x_i = 1$) or not ($x_i = 0$). However, in practice we want to know the asset weights. To introduce real weights instead of binary flags and at the same time preserve the binary nature of the optimization, we follow the suggestion of \cite{ptf-optim-dynamic} and define the weight of the $i$\textsuperscript{th} asset as a \textit{binary fraction} $w_i = \sum_{k=1}^{\ell} 2^{-k}x_{i}^{(k)}$. Parameter $\ell$ is the number of bits dedicated to expressing the weight and $x_{i}^{(k)}$ is a binary variable containing the $k$\textsuperscript{th} decimal place of the weight. 

Replacing the binary flags $x_i$ in \eqref{eq_qubo_markowitz_bin} with weights $w_i$ leads to the Markowitz portfolio optimization with real asset weights\footnote{In fact, the weights are not real but only rational because of the finite number of binary variables expressing them. However, this is also the case with ``real'' numbers in classical computing, as memory is always finite.}
\begin{equation}
\label{eq_qubo_markowitz_weight}
-\sum_{i=1}^n r_i \sum_{k=1}^{\ell} 2^{-k}x_{i}^{(k)} 
+ \lambda\sum_{i=1}^n\sum_{j=1}^n c_{ij} \Big(\sum_{k=1}^{\ell} 2^{-k}x_{i}^{(k)}\Big) \Big(\sum_{k=1}^{\ell} 2^{-k}x_{j}^{(k)}\Big)
+ F\Big(\sum_{i=1}^n p_i \sum_{k=1}^{\ell} 2^{-k}x_{i}^{(k)} - B\Big)^2 
\rightarrow  \min.
\end{equation}
Note that we ignored the sign of weight $w_i$. Introducing the sign is possible,\footnote{\cite{qubo-regression} advise rewriting the weight  as $w_i = \sum_{k=1}^{\ell} 2^{-k}x_{i}^{(k),\text{pos}} - \sum_{k=1}^{\ell} 2^{-k}x_{i}^{(k),\text{neg}}$.} but we will assume that short positions are forbidden, hence we can assume that the weights are non-negative. Thanks to the approach used to binarize the weights, it holds that $w_i \in \langle 0;1 \rangle \,\, \forall i$, which is the set of constraints used in the Markowitz model.

To make the problem independent of the budget volume $B$, we set $B = 1$ and $p_i = 1 \,\,\forall i$, which changes the budget constraint to $\sum_i w_i = 1$. This is another constraint generally used in Markowitz portfolio optimization. Problem \eqref{eq_qubo_markowitz_weight} also ignores the time dimension of the portfolio optimization. It is common in practice to rebalance the portfolio periodically because of changes in a benchmark, or to radically change the portfolio composition during or after major crises. This can be reflected in the optimization by introducing a time-dependent covariance matrix and returns and adding other binary variables containing asset weights for time $t=1,2 \dots T$. What is more, any change in the portfolio structure leads to transaction costs, which we naturally want to minimize. The costs are proportional to the changes in asset weights between time periods $t-1$ and $t$. We therefore introduce variables $x_i^{(k,t)}$ expressing the $k$\textsuperscript{th} decimal place of the $i$\textsuperscript{th} asset weight at time $t$. Note that $x_i^{(k,0)} = 0 \,\,\ \forall i,k$.\footnote{This means that the portfolio is composed of cash only at time $t=0$.} The costs incurred in connection with changing the position in the $i$\textsuperscript{th} asset are $\nu_i^{(t)} |\sum_{k=1}^\ell x_i^{(k,t)}-x_i^{(k,t-1)}|$, where $\nu_i^{(t)}$  are unit transaction costs. To preserve the QUBO nature of our optimization problem, we replace the absolute value with the square. After the inclusion of independence of the absolute budget volume, the time dimension and transaction costs, problem \eqref{eq_qubo_markowitz_weight} is transformed to
\begin{equation}
\label{eq_qubo_markowitz_dynamic}
\begin{aligned}
\sum_{t=1}^T \Big[ & \, &  \text{time dimension} \\ 
\,  &  -\sum_{i=1}^n r_i^{(t)} \sum_{k=1}^{\ell} 2^{-k}x_{i}^{(k,t)}  & \text{return} \\
\,  & + \lambda\sum_{i=1}^n\sum_{j=1}^n c_{ij}^{(t)} \Big(\sum_{k=1}^{\ell} 2^{-k}x_{i}^{(k,t)}\Big) \Big(\sum_{k=1}^{\ell} 2^{-k}x_{j}^{(k,t)}\Big)  & \text{risk}\\
\,  & + \mu\sum_{i=1}^n \nu_i^{(t)} \Big[\sum_{k=1}^{\ell} 2^{-k}\big(x_{i}^{(k,t)} - x_{i}^{(k,t-1)}\big)\Big]^2  & \text{transaction costs} \\
\,  & + F\Big(\sum_{i=1}^n\sum_{k=1}^{\ell} 2^{-k}x_{i}^{(k,t)} - 1\Big)^2  & \text{sum of weights equal 1}\\
\Big] \rightarrow  \min. & \, & \,
\end{aligned}
\end{equation} 

Note that parameter $\mu \ge 0$ is the sensitivity of a portfolio manager to the transaction costs. Problem \eqref{eq_qubo_markowitz_dynamic} is the final version of the \textit{discrete-time dynamic unconstrained binary portfolio optimization} based on the Markowitz model. Note that the more convenient matrix form of \eqref{eq_qubo_markowitz_dynamic} is provided in Appendix~\ref{appendixMatrixObj}.

\subsubsection*{Computational Complexity}

We now turn our attention to the computational complexity of the problem introduced above. QUBO is an exponentially complex problem on a classical computer, as adding one binary variable doubles the possible inputs of the objective function.\footnote{We use the term \textit{exponentially complex} rather loosely. The precise statement should be that the problem belongs to the $\mathbf{NP}$ complexity class. By stating that the problem is exponentially complex, we mean that currently there is no known algorithm capable of solving the problem in polynomial time. But this does not exclude the possibility of such an algorithm being discovered in the future, as whether $\mathbf{P} = \mathbf{NP}$ or $\mathbf{P} \ne \mathbf{NP}$ remains unresolved. Note that we will use this simplification in the rest of the paper to reflect the current capabilities of computational tools. Readers interested in learning more about complexity theory can consult \cite{misc-nielsen-chuang-book}, chapter 3.} It may seem that by moving from the continuous to the QUBO version, we have made the problem harder, but any general problem of quadratic programming with real variables is exponentially complex on a classical computer, as proved by \cite{misc-qp-np} and \cite{misc-qp-np2}. On the other hand, if matrix $\mathbf{A}$ in the objective function \eqref{eq_qubo_matrix} is positive definite and the constraints are convex, the quadratic program can be solved in polynomial time, as shown by \cite{misc-qp-convex-orig} and \cite{misc-qp-convex}. 

A covariance matrix is positive definite. However, adding a general penalty function can lead to an objective function with an indefinite matrix. Especially in this most difficult version of the problem, quantum algorithms could offer speed-up. What is more, if we were able to show speed-up even for cases that a classical computer can solve in polynomial time (such as the one we present in this paper), we would have stronger empirical proof of better performance of quantum algorithms in some cases of quadratic programming and we would therefore show that they are useful in portfolio optimization in general.

\section{Algorithms for QUBO}
\label{secAlg}
This part is devoted to describing algorithms for solving QUBO problems. First, we will discuss the classical algorithms which we compare the quantum ones with. After that, we will describe the QAOA and VQE quantum variational algorithms intended for a universal gate-based quantum computer. We will then describe D-Wave's quantum annealer, the physical realization of the adiabatic approach. Finally, a QUBO algorithm based on Grover database quantum searching will be discussed.

\subsection{Classical Algorithms}
\label{subsecAlgClassical}
In this subsection we will present classical methods for quadratic optimization. The first one, the gradient method, is intended for solving continuous problems, because the original Markowitz model works with continuous variables. The method will serve as a benchmark allowing us to decide whether the transformation of Markowitz continuous portfolio optimization to QUBO and the application of quantum computers confers any advantage.

As noted in the previous part, QUBO is an exponentially complex problem on a classical computer. This renders an exhaustive search  for the solution (the brute force method) impossible. Therefore, more sophisticated algorithms have been devised. First, we will describe the branch and bound algorithm, which is an exact solver for integer programming. Then, we will discuss two optimization heuristics, namely simulated annealing and genetic optimization. Note that unlike the branch and bound method, these heuristics are only able to find sub-optimal solutions, as they are based on random searching in the problem domain.

Besides the above-mentioned algorithms, there is a plethora of other QUBO heuristics and exact solvers. Interested readers can consult the overview by \cite{classical-qubo}. There are also several software packages intended for solving optimization problems including QUBO, for example CPLEX\textsuperscript{TM} provided by \cite{classical-cplex} and XPRess\textsuperscript{TM} offered by \cite{classical-xpress}. An open source alternative is BiqCrunch designed by \cite{classical-biq-crunch}.

\subsubsection*{Gradient Method}

As noted above, we include gradient optimization among the classical methods for comparing the QUBO approach to portfolio optimization with continuous quadratic programming. For our purposes we will use the implementation of the method in the MS~Excel solver. We will not discuss the method further, because it is widely known and long established.\footnote{The method was proposed by the famous French mathematician Louis Augustin Cauchy back in 1847.} Interested readers can find details in \cite{classical-gradient}.

\subsubsection*{Branch and Bound Algorithm}

This algorithm was proposed by \cite{classical-branch-bound} and was originally intended for integer optimization. As binary optimization is a special case of integer optimization, the algorithm can be used for QUBO as well.\footnote{See the application of the branch and bound algorithm to the travelling salesman problem by \cite{classical-branch-tsp}.}

The main idea behind the algorithm is to cut off parts of a problem domain that offer no improvement of the objective function value found so far. First, the problem is solved with no integer constraints imposed on the variables. Once the solution is reached, the problem domain is divided into two subsets and a special constraint for some variable $x_i$ is added for each subset. Assume that $a$ is an integer part of the value of variable $x_i$. Then constraint $x_i \le a$ is added for the first subset and  $x_i \ge a+1$ for the second one. A similar approach can be employed in the case of binary optimization. However, the constraints are simpler: constraint $x_i = 0$ is added for one subset and constraint $x_i = 1$  for the other. The lower bound of the objective function is calculated  for each subset, and the subset with the higher bound is no longer taken into consideration. This effectively reduces the size of the domain which has to be searched for the optimal solution. The process is repeated until the optimal solution is found.

Note that the branch and bound algorithm can degenerate into the brute force method. This means that in the worst case it exhibits exponential complexity, as discussed by \cite{classical-branch-complexity}. The algorithm is exact, that is, it guarantees that the optimal solution will be always found, but, as already mentioned, sometimes at the cost of a long run time.

We will use the implementation of the algorithm in the CPLEX\textsuperscript{TM} software package provided by \cite{classical-cplex}. The product has a free version with the number of variables and constraints limited to 1,000. Nevertheless, this is sufficient for our purposes.

\subsubsection*{Simulated Annealing}

The first heuristic intended for QUBO, simulated annealing, is inspired by the cooling of a hot material. The algorithm was first proposed by \cite{classical-annealing-metropolis} for finding thermal equilibrium in chemical processes. Later, it was improved by \cite{classical-annealing-continuous} for the optimization of general continuous functions. Many special-purpose adaptations of the algorithm have been designed, for example for simulations in statistical physics \cite{classical-annealing-coulomb} and for portfolio optimization \cite{classical-annealing-portfolio}. For some problems with an objective function defined on discrete domains, including binary optimization problems, the investigation carried out by \cite{classical-annealing-discrete} revealed superior behaviour of the algorithm in comparison with other heuristics.

The principle of simulated annealing can be briefly described as follows. As a material cools down, the particles the material is composed of find new positions so that the total energy stored in the material is the lowest possible. The configuration can be likened to a bit string composed of binary variables and the energy to the respective value of the objective function. The new configuration is found by means of random changes in the configuration of the particles. To avoid getting stuck in local minima of the objective function, a configuration with higher energy (i.e.~a worse solution) is sometimes allowed.\footnote{This is a natural process occurring in any matter caused by fluctuations at the atomic level.} The particular implementation of the random changes (known as the \textit{noise operator}), the \textit{cooling schedule} (defining the speed of cooling) and the rule for temporary acceptance of a higher energy configuration can be various. We will use our own implementation of the simulated annealing algorithm based on the studies listed above and adapted to binary optimization. Technical details of the implementation are available in Appendix~\ref{appendixSrcSimAnneal}, and the source code is provided by \cite{classical-sa-mtl-implement}.

\subsubsection*{Genetic Optimization}

The second heuristic, genetic optimization, is based on the simulation of an evolutionary process. Interestingly, the idea of solving computational problems with a biologically inspired system appeared in \cite{phil-turing}. He was trying to contradict the statement that a calculation machine cannot think, as expressed by \cite{phil-lovelace}, among others, and proposed a computational model based on the mind of a child trying to find a solution to a problem by trial and error, similarly to the evolutionary process. Evolution-based algorithms were investigated more seriously by \cite{classical-genetic-first1} and \cite{classical-genetic-first2}. Since then, many genetic algorithms have been proposed. A vast overview of these techniques is provided by \cite{classical-genetic-overview}. 

In a nutshell, the genetic approach works as follows. The bit string representing the argument of the objective function is considered to be the genome of an organism. The objective function value assigned to the argument is a measure of the \textit{fitness} of the organism. The goal of genetic optimization is to modify the genome so as to achieve the best possible fitness. To do so, \textit{mutation and crossing operators} are introduced. The mutation operator randomly changes the genome; for example, it flips a randomly chosen binary variable(s). The crossing operator picks up two or more organisms from the \textit{population} (i.e.~ the set of bit strings with ``good fitness'') according to a predefined scheme and returns a combination thereof. If the fitness of the newly produced organism is better than that of the organisms ``created'' so far, it becomes a member of the population. Organisms with worse fitness are removed from the population. The exact definitions of the mutation and crossing operators and the rules for adding and removing organisms from the population depend on the particular genetic algorithm. We will use the implementation of the algorithm provided by \cite{classical-genetic-mtl-implement}.

\subsection{Quantum Adiabatic Approach}
\label{subsecAlgAdiabatic}
The quantum computer was originally proposed by \cite{misc-feynman-intro} as a tool for simulating quantum systems, a problem that is exponentially complex on a classical computer. Later, Feynman's proposal was investigated by \cite{misc-loyd-conjecture}, who conjectured that quantum systems could in some cases be simulated in polynomial time on a quantum computer. Although Lloyd's conjecture remains unproven,\footnote{\cite{misc-sampling-polynomial} showed that random circuit sampling, a problem for which Google has claimed a quantum advantage and which until recently was considered to be exponentially complex, can be solved under specific circumstances in polynomial time, although the polynomial is of high degree. This may increase the doubts about Lloyd's conjecture, but further research is still necessary.} if we can figure out how to convert an optimization problem (QUBO in our case) into the simulation of a quantum system, we can exploit the potential power of quantum computers (see the list of articles in the Introduction providing empirical evidence that quantum computers can solve some optimization problems faster than their classical counterparts).

To carry out a quantum system simulation, the quantum computer has to get a description of the system. In quantum physics, such description is the \textit{Hamiltonian} (or energy operator).  It characterizes the allowed energy levels of the system and the energy the system takes on at a particular level. Among energy levels, a special place is dedicated to the \textit{ground state}, the state with the lowest possible energy and associated with the optimal solution of the optimization problem. In mathematical terms, the Hamiltonian is generally a Hermitian operator. However, in the following text, we will assume the Hamiltonian to be a Hermitian matrix $\mathcal{H}$. The energy of the quantum system described by $\mathcal{H}$ being in state $|\psi\rangle$ is given by the expression $E=\langle\psi|\mathcal{H}|\psi\rangle$.\footnote{For more information on Hamiltonians, see \cite{misc-nielsen-chuang-book}, chapter~2.2.2. For details concerning the simulation of quantum systems, see \cite{misc-hamiltonian-simulation}, chapter~9.}

A special kind of Hamiltonian is the \textit{Ising Hamiltonian}, defined as
\begin{equation}
\label{alg_adiabatic_ising}
\mathcal{H}_{\text{Ising}} = \sum_{i=1}^{n}\sum_{j=1}^{n}Q_{ij}\mathbf{Z}_i \otimes \mathbf{Z}_j + \sum_{i=1}^n c_i \mathbf{Z}_i, 
\end{equation}
where $\mathbf{Z}_i$ denotes a quantum gate composed of a $\mathbf{Z}$ gate applied to the $i$\textsuperscript{th} qubit and identity operators applied to the other qubits, and term $\mathbf{Z}_i \otimes \mathbf{Z}_j$ means a quantum gate composed of two $\mathbf{Z}$ gates applied to the $i$\textsuperscript{th} and $j$\textsuperscript{th} qubits while identity operators are applied to the other qubits and $Q_{ij}, c_i \in \mathbb{R}$.\footnote{A more precise term than gate would be operator, because Hamiltonian simulation can be carried out on a general quantum computer, not only on a gate-based machine. However, despite the name, a $\mathbf{Z}$ operator and a $\mathbf{Z}$ gate are mathematically the same object.} There is a clear resemblance between the Ising Hamiltonian \eqref{alg_adiabatic_ising} and the general QUBO problem \eqref{eq_qubo_sum}. The ground state of the Ising Hamiltonian \eqref{alg_adiabatic_ising} can be identified with the optimal solution of problem \eqref{eq_qubo_sum}. This means that we have translated the QUBO problem to the simulation of a quantum system. Technical details of the translation, mainly the relations between the coefficients and variables in the QUBO problem \eqref{eq_qubo_sum} and the Ising Hamiltonian \eqref{alg_adiabatic_ising}, are provided by \cite{qubo-ising-formulations}.

The next step is to simulate the Ising Hamiltonian and search for its ground state.\footnote{\cite{misc-loyd-conjecture} included a system described with Ising Hamiltonians among examples of quantum systems that could be simulated on a quantum computer with exponential speed-up if his conjecture is confirmed.} To do so, we leverage the \textit{adiabatic theorem}. Assume that the quantum system is described by the initial Hamiltonian $\mathcal{H}_0$ at time $t = 0$. At the same time, the system is in the ground state of $\mathcal{H}_0$. Subsequently, the Hamiltonian of the system is changed. At time $t \in \langle 0;1\rangle$, the system is described by the Hamiltonian 
\begin{equation}
\mathcal{H}(t) = (1-t)\mathcal{H}_0 + t\mathcal{H}_1. 
\end{equation}
If the change is slow enough, the system remains in the ground state of $\mathcal{H}(t)$ for any $t$. This means that finally (i.e.~at time $t=1$) the system is in the ground state of $\mathcal{H}_1$. In our case, we set $\mathcal{H}_1 = \mathcal{H}_\text{Ising}$.

Let's turn our attention to Hamiltonian  $\mathcal{H}_0$. If $\mathcal{H}_0$ is simple enough, we are able to determine its ground state analytically. Such simple Hamiltonian is 
\begin{equation}
\mathcal{H}_0 = \sum_{i=1}^n \mathbf{X}_i, 
\end{equation}
where $\mathbf{X}_i$ denotes a quantum gate composed of an $\mathbf{X}$ gate applied to the $i$\textsuperscript{th} qubit and identity operators applied to the other qubits. The ground state of Hamiltonian $\mathcal{H}_0$ is the uniformly distributed superposition $\frac{1}{\sqrt{2^n}}\sum_{i=0}^{2^n-1}|i\rangle$, where $|i\rangle$ is the $i$\textsuperscript{th} basis state of a quantum system composed of $n$ qubits. Knowing the ground state of $\mathcal{H}_0$ and employing the adiabatic theorem enables us to find out the ground state of $\mathcal{H}_\text{Ising}$ associated with the optimal solution of the QUBO problem. For a better understanding of how simulation based on the adiabatic theorem works, see the analogy in Box~1.

\subsection{Variational Algorithms and Quantum Annealers}
\label{subsecAlgVariational}
The previous section provided the theoretical framework for the simulation of Ising Hamiltonians and equivalent QUBO problems. However, this approach has to be implemented on a real quantum computer, which we will discuss in this part. First, we will present algorithms intended for a gate-based universal quantum computer (\textit{variational algorithms}) and then we will turn our attention to a single-purpose \textit{quantum annealer}.

\vspace{0.2cm}
\noindent
\fbox{%
\parbox{\textwidth}{
\textbf{Box 1: Adiabatic Theorem Analogy}
\vspace{0.1cm}

The adiabatic theorem can be explained with an analogy. Assume we have a sheet of paper scattered with iron filings. This is our system described by the initial Hamiltonian $\mathcal{H}_0$. If no external force is applied, the system remains unchanged. Imagine that we slowly move a magnet under the sheet. The filings begin to move and follow the magnetic lines of force. In the end, we will see a typical pattern on the sheet -- the iron filings oriented in the direction of the external magnetic field. This is the system described by the final Hamiltonian $\mathcal{H}_1$. The systems that exist when the magnet is only partially under the sheet are described by Hamiltonians $\mathcal{H}(t)$. We now turn our attention to the requirement of carrying out the changes slowly. If we move the magnet quickly, some of the filings remain stuck because of friction between them and the sheet. Clearly, these filings resist the magnetic field and they have to have enough energy to do so. As additional energy is needed, the system is clearly not in its ground state. However, slow changes allow the filings to adapt to the increasing magnetic force without getting stuck, hence they need less energy. In other words, the system remains in the ground state all the time.
}}

\subsubsection*{Variational Algorithms}

\cite{misc-adiab-gate-equivalent} showed that the adiabatic and universal gate-based quantum computing models are equivalent.\footnote{To be precise, \cite{misc-adiab-gate-equivalent} stated that \textit{the model of adiabatic computation is polynomially equivalent to the standard model of quantum computation.} On the other hand, \cite{qubo-preskill} pointed out that this conclusion is valid only for noiseless qubits, and at the cost of adding a high number of ancilla qubits despite the polynomial equivalence.} This means that algorithms can be designed for a gate-based quantum computer implementing the above-described simulation of Ising Hamiltonians. What is more, the equivalence of the models allows us to expect that algorithms based on the adiabatic model will provide exponential speed-up if the conjecture by \cite{misc-loyd-conjecture} is confirmed. At present, we have to assume that those algorithms offer unproven speed-up, but, as discussed in the Introduction, they show promising results in terms of improved computational time for some specific problems. In what follows, we describe two algorithms intended for solving QUBO problems and developed for gate-based quantum computers.

The first is the Quantum Approximate Optimization Algorithm (QAOA) introduced by \cite{qaoa-orig-paper} and improved for some kinds of QUBO problems (including portfolio optimization) by \cite{qaoa-warm-start}. The second is the Variational Quantum Eigensolver (VQE) developed by \cite{vqe-orig-paper} and improved by \cite{vqe-improved}. Both algorithms are \textit{hybrid}, which means that they combine the classical and quantum approaches. While the evolution of the quantum system state (the hard part) runs on a quantum computer, the calculation of the energy of the quantum state (the easy part) runs on a classical computer. In the first step, a circuit describing the Hamiltonian $\mathcal{H}(t)$ is constructed. Because of parameter $t$, the circuit has to be parametric as well. Moreover, we have to ensure that the simulation runs slowly to preserve the assumption of the adiabatic theorem. This all means that we have to run the simulation several times with different parameters and find the circuit describing the simulation of $\mathcal{H}(t)$ in the best possible way. In other words, we solve the classical optimization problem $E=\langle\psi(\mathbf{\theta})|\mathcal{H}_\text{Ising}|\psi(\mathbf{\theta})\rangle \rightarrow \min_\mathbb{\mathbf{\theta}}$, where $\mathbf{\theta}$ is a vector of parameters figuring in the quantum circuit. Since an Ising Hamiltonian comprises only two-body interactions, there is a technique described by \cite{vqe-improved} allowing us to calculate $E$ efficiently on a classical computer. The result of the optimization is the vector $\mathbf{\theta}_\text{optim}$ allowing us to construct a circuit for finding the ground state of $\mathcal{H}_\text{Ising}$.  

\begin{figure}[hbt]
\caption{QAOA Single-Layer Circuit for Two-Variable QUBO Problem}
\begin{center}
\includegraphics[scale = 0.9]{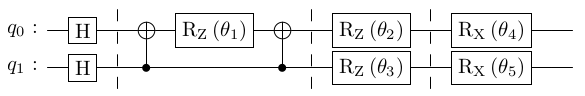}
\end{center}
\vspace{0.1cm}
{\footnotesize \textbf{\textit{Note:}} Qubits $q_0$ and $q_1$ represent binary variables $x_0$ and $x_1$ respectively. The Hadamard gates $\mathbf{H}$ prepare a uniformly distributed quantum state. Then the Ising Hamiltonian is applied: the $\mathbf{Rz}$ gate on qubit $q_0$ together with the two CNOT gates represent a quadratic term $x_0x_1$, and the other $\mathbf{Rz}$ gates are connected with linear terms $x_0$ and $x_1$. The $\mathbf{Rx}$ gates are related to the initial Hamiltonian $\mathcal{H}_0$. Note that the $\mathbf{Rx}$ and $\mathbf{Rz}$ gates are the exponentials of the $\mathbf{X}$ and $\mathbf{Z}$ gates figuring in the initial and Ising Hamiltonians respectively. Coefficients $Q_ij$ and $c_i$ in the Ising Hamiltonian and parameters $\beta_j$ and $\gamma_j$ are hidden in the rotational angles of the $\mathbf{Rx}$ and $\mathbf{Rz}$ gates, i.e.~$\theta_1 \dots \theta_5$.}\\
{\footnotesize \textbf{\textit{Source:}} Adapted from the result obtained in the IBM Q\textsuperscript{TM} environment}
\label{fig_alg_adiabatic_qaoa}
\end{figure}

Both the QAOA and the VQE are based on the approach described above, but they differ in the circuit employed for finding the ground state. The QAOA uses a circuit described with the quantum gate 
\begin{equation}
\label{alg_adiabatic_qaoa}
\mathbf{U}(\beta, \gamma) = \prod_{j=1}^{p} \mathrm{e}^{-i\beta_j \mathcal{H}_0} \mathrm{e}^{-i\gamma_j \mathcal{H}_\text{Ising}},
\end{equation}
where $p$ is the number of circuit layers and $\beta_j$ and $\gamma_j$ are parameters connected with the $j$\textsuperscript{th} layer of the circuit. Number $p$ is a user-defined value. Increasing $p$ leads to a higher probability of finding the optimal solution, but also increases the complexity of the circuit. For $p \rightarrow \infty$, finding the optimal solution is guaranteed, but for finite $p$, the algorithm is only a heuristic. An example of a QAOA circuit is provided in Figure~\ref{fig_alg_adiabatic_qaoa}.

\begin{figure}[hbt]
\caption{VQE Double-Layer Circuit for Three-Variable QUBO Problem}
\begin{center}
\includegraphics[scale = 0.8]{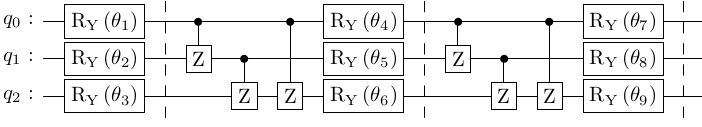}
\end{center}
\vspace{0.1cm}
{\footnotesize \textbf{\textit{Note:}} The circuit has two parts -- an initial layer composed only of $\mathbf{Ry}$ gates and at least one layer composed of $\mathbf{Ry}$ gates and controlled $\mathbf{Z}$ gates. The latter interconnects the qubits and enables the VQE to take the quadratic terms in the QUBO problem into account. The rotational angles of the $\mathbf{Ry}$ gates are optimized so that the circuit simulates the quantum system described by the Ising Hamiltonian and finds its ground state.}\\
{\footnotesize \textbf{\textit{Source:}} Adapted from \cite{vqe-improved}}
\label{fig_alg_adiabatic_vqe}
\end{figure}

As can be seen from equation~\eqref{alg_adiabatic_qaoa} and Figure~\ref{fig_alg_adiabatic_qaoa}, the structure of the Hamiltonian $\mathcal{H}(t)$ is reflected in the quantum circuit. In contrast, the composition of the VQE circuit is very different. The circuit is built from several layers of controlled $\mathbf{Z}$ gates and $\mathbf{Ry}$ gates. The rotational angles of $\mathbf{Ry}$ serve as the optimized parameters $\mathbf{\theta}$. The structure of the Ising Hamiltonian is not reflected in the circuit. Because of this, the VQE is more general, and the QAOA can be considered a special case of it. An example of a VQE circuit is provided in Figure~\ref{fig_alg_adiabatic_vqe}.\footnote{The VQE circuit is often called \textit{ansatz} (an educated guess), because after proper parametrization it can be used for simulating a quantum system and finding its ground state.}

\subsubsection*{Quantum Annealers}

Besides variational algorithms for universal gate-based quantum computers, single-purpose \textit{quantum annealers} exploiting the adiabatic approach have been devised. In this part we will focus on the annealers provided by D-Wave.

D-Wave annealers are in fact a quantum processor based on superconducting qubits.\footnote{See \cite{misc-superconducting-processors} for more on superconducting processors.} For each qubit, it is possible to set the probability amplitudes of states $|0\rangle$ and $|1\rangle$ -- the \textit{bias}. Pairs of qubits can be connected together with \textit{couplers} to prepare an entangled state.\footnote{Note that because of the special topology of the processor, there is no all-to-all connectivity among the qubits. As a result, more than two qubits can be involved in the quadratic terms of an objective function.} The probability amplitudes of the basis states that the entangled state is composed of can be set by manipulating the coupler \textit{strength}. Setting the biases and strengths enables one to programme a Hamiltonian into the annealer. The annealer function is again based on the adiabatic theorem. First, the initial Hamiltonian $\mathcal{H}_0$ is programmed into the annealer, which is then set to be in state $\frac{1}{\sqrt{2^n}}\sum_{i=0}^{2^n-1}|i\rangle$. In the next step, the biases and strengths are slowly changed so that the annealer behaves according to the Hamiltonian $\mathcal{H}(t)$. At the same time, it remains in the ground state of $\mathcal{H}(t)$. In the end, only the Ising Hamiltonian describes the annealer behaviour. Finally, a measurement is carried out to get the ground state of the Ising Hamiltonian (i.e.~the solution of the QUBO problem). 

As with any other quantum computation, superposition and entanglement are involved in the described process. On top of that, the phenomenon of \textit{quantum tunnelling} plays a role as well. Quantum tunnelling allows the annealer to get out of local minima -- ``valleys'' -- by tunnelling through the ``walls'' of the valley (see Box~2 for more about quantum tunnelling). 

\vspace{0.2cm}
\noindent
\fbox{%
\parbox{\textwidth}{
\textbf{\textit{Box 2: Quantum Tunnelling}}
\vspace{0.1cm}

In the classical world, it is impossible for a particle to go through an energy barrier. To overcome the barrier, the particle has to gain a sufficient amount of energy from an outside source. However, in the quantum world, the particle can ``borrow'' the energy $\Delta E$ for time $\Delta t$. It holds that the higher the energy, the shorter the time for which the energy can be ``borrowed''. With the ``borrowed'' energy, the particle overcomes the barrier and then the energy is ``redeemed''. This process is a consequence of Heisenberg's relations of uncertainty. In this particular case, the second relation $\Delta E \Delta t \ge h/(4\pi)$ plays a role. Note that $h$ is the Planck constant ($h = 6.6261 \times 10^{-34} m^2 kg s ^{-1}$).
}}
\vspace{0.2cm}

It is worth noting that due to quantum fluctuations the annealer does not have to remain in the ground state of the Hamiltonian  $\mathcal{H}(t)$ all the time. As a result, the solution is only a sub-optimal one and quantum annealers have to be considered a heuristic tool similar to the QAOA and the VQE.

D-Wave also offers \textit{hybrid heuristics}. This approach combines classical heuristic algorithms with quantum annealing. For example, the output of classical heuristics is used as the initial solution for the quantum annealer. The problem can also be broken down into sub-problems and each part solved with a different algorithm -- classical or quantum. Finally, the sub-problem results are combined to get the solution of the complete problem. D-Wave offers hybrid heuristics for both unconstrained and constrained optimization.\footnote{In our case, the constrained heuristic allows constraint $\sum_i w_i^{(t)} = 1 \,\, \forall t$ to be expressed outside the objective function.}

\cite{dwave-manual} provides a detailed description of quantum annealers. Further information on D-Wave hybrid heuristics is presented in \cite{dwave-hybrid}. Complete documentation for the D-Wave Leap\textsuperscript{TM} and Ocean\textsuperscript{TM} development environments is provided in \cite{dwave-ocean}. The source codes used in our tests are available at GitHub \cite{dwave-srccodes}.

\subsection{Grover Adaptive Search}
\label{subsecAlgGrover}
In this section, we will discuss the QUBO method based on the algorithm designed by \cite{grover-orig-paper}. Originally, the algorithm was intended for fast search in an unordered database, i.e.~a database without an index or a special structure. As we have no index, with a classical computer we have to go through the database records one by one and check whether the record currently picked out is the one we are searching for. In the worst case, the searched record is the last one. Clearly, the problem has linear complexity $O(N)$, where $N$ is the number of records. With Grover quantum search, the complexity drops to $O(\sqrt{N})$, i.e.~the algorithm provides quadratic speed-up. Unfortunately, the practicality of the Grover algorithm for database searching was questioned by \cite{grover-practicality}. 

Notwithstanding the practical limitation connected with the original purpose of the algorithm, based on ideas presented by \cite{grover-fmin-ideas}, the Grover algorithm was modified to serve as a QUBO algorithm by \cite{grover-qubo}. In contrast to adiabatic approaches, the Grover algorithm offers proven quadratic speed-up. The complexity of QUBO is exponential ($O(2^n)$, $n$ being the number of variables) on a classical computer. The Grover algorithm thus reduces the complexity to $O(2^{n/2})$, which is still exponential in $n$, but any improvement helps. However, it is important to note that in the current era of noisy quantum computers, the Grover algorithm is not able to provide the declared quadratic speed-up. As shown by \cite{misc-nisq-complexity}, its complexity remains the same as in the case of a classical unstructured search algorithm.\footnote{Interestingly, this holds for any algorithm offering quadratic speed-up, including quantum Monte Carlo.}

\subsubsection*{Grover Algorithm in General}

\begin{figure}[H]
\caption{Illustration of Grover Algorithm Action}
\begin{center}
\includegraphics[scale = 0.55]{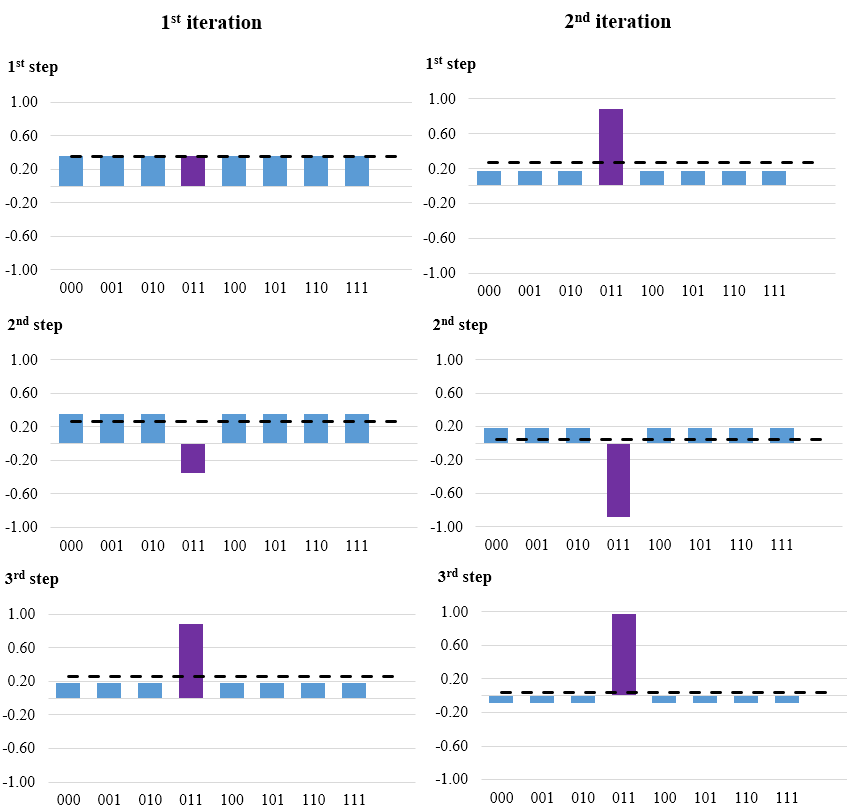}
\end{center}
\vspace{0.1cm}
{\footnotesize \textbf{\textit{Note:}} Assume that we want to find a record identified with state $|011\rangle$ (highlighted in violet). In the first iteration, a uniformly distributed superposition of all the records is prepared (i.e.~the probability of measuring any basis state is 12.5\%, or the probability amplitude is 0.3536). Next, the probability amplitude of state $|011\rangle$ is inverted. This leads to a decrease in the average probability amplitude (depicted with a dashed black line). After that, the probability amplitudes are ``mirrored'' around the average. The quantum state gained is inputted into the second iteration. Then marking and ``mirroring'' is carried out again. As a result, the probability amplitude of state $|011\rangle$  is close to 1. We found the desired record after two iterations. On a classical computer, the search could take up to eight iterations.}\\
{\footnotesize \textbf{\textit{Source:}} Author's own creation}
\label{fig_grover_amplitudes}
\end{figure}

First, we will outline how the original Grover algorithm works. Assume that a database contains $N = 2^n$ records, where $n$ is the number of qubits used for representation of the database records. First of all, superposition $\frac{1}{\sqrt{2^n}}\sum_{i=0}^{2^n-1}|i\rangle$, containing all the database records with the same probability, is prepared. The next step is to mark the basis state representing the record we are searching for by inverting its probability amplitude. As a result, the average probability amplitude of the quantum state decreases. After that, all the probability amplitudes are ``mirrored'' around the average amplitude. This leads to an increase in the probability amplitude of the marked state and a decrease in the other amplitudes. The marking and ``mirroring'' steps are repeated until the probability amplitude of the marked state is not close to 1. Finally, the qubits are measured to reveal the marked state, or in other words to return the database record we were searching for. Note that for a database with $N$ records, we have to repeat the marking and ``mirroring'' $R =\lceil\pi \sqrt{N}/4\rceil$ times. The described process is graphically illustrated in Figure~\ref{fig_grover_amplitudes} on a search for state $|011\rangle$ in a database containing eight records. 

\begin{figure}[h]
\caption{Grover Algorithm Circuit}
\begin{center}
\includegraphics[scale = 0.7]{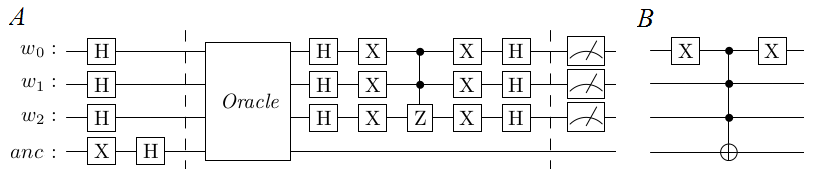}
\end{center}
\vspace{0.1cm}
{\footnotesize \textbf{\textit{Note:}} Panel A depicts a general Grover algorithm circuit for three working qubits. The part of the circuit between the dashed lines (i.e.~the oracle and the diffusion operator) is repeated $R$ times. Note that $w$ stands for working qubits and $anc$ for ancilla qubit. Panel B shows the oracle for marking state $|011\rangle$. For a higher number of qubits, the circuit can be scaled up following the patterns in the initial phase and the diffusion operator. Interested readers can learn more about the circuit construction using the interactive applet created by \cite{grover-simulator}.}\\
{\footnotesize \textbf{\textit{Source:}} Adapted from \cite{grover-simulator}}
\label{fig_grover_circuit}
\end{figure}

We now turn our attention to the technical realization of the Grover algorithm. The circuit implementing the algorithm is depicted in Figure~\ref{fig_grover_circuit}. Without loss of generality, we will assume that we are searching for record $|011\rangle$ in a database stored within three qubits. We refer to those qubits as \textit{working qubits}. At the beginning, the working qubits are set to be in equally distributed superposition with Hadamard gates. Regardless of the number of working qubits, one \textit{ancilla qubit} is added. The ancilla is initialized to state $|-\rangle$. This means that at the beginning, the quantum computer is in state 
\begin{equation}
|\psi_\text{init}\rangle = \Big(\frac{1}{\sqrt{8}}\sum_{i=0}^7|i\rangle\Big) \otimes |-\rangle. 
\end{equation}

After that, a so-called \textit{oracle} is applied to mark the record we are searching for. In fact, the oracle is a multi-qubit controlled $\mathbf{X}$ gate negating the target qubit if a logical expression based on the control qubits values is true. In our example, the oracle negates the ancilla if the working qubits are in state $|011\rangle$, otherwise the ancilla remains unchanged. Negation of $|-\rangle$ leads to state $-|-\rangle$, hence state $|\psi_\text{init}\rangle$ is changed to\footnote{It holds that $\mathbf{X}|-\rangle = \frac{1}{\sqrt{2}}(\mathbf{X}|0\rangle - \mathbf{X}|1\rangle) = \frac{1}{\sqrt{2}}(|1\rangle - |0\rangle) = -\frac{1}{\sqrt{2}}(|0\rangle - |1\rangle) = -|-\rangle$. It may seem that the minus sign before the state can be ignored, as it is a global quantum phase. However, state $|-\rangle$ is incorporated into multi-qubit state   $|\psi_\text{init}\rangle$ and the sign is changed only for state $|011\rangle$, hence -1 is in fact the relative phase.}
\begin{equation}
|\psi_\text{oracle}\rangle = \Big(\frac{1}{\sqrt{8}}\sum_{i \in \{0,1,2,4,5,6,7\}}|i\rangle \textcolor{red}{- \frac{1}{\sqrt{8}}|011\rangle}\Big) \otimes |-\rangle.
\end{equation}

Next, a so-called \textit{diffusion operator} implementing the ``mirroring'' is applied. The logic behind the construction of the diffusion operator is beyond the scope of this article. However, details are provided by \cite{grover-idea-behind}.\footnote{The article contains ideas leading to the construction of the Grover algorithm. Interestingly, the algorithm is a by-product of Schr\"{o}dinger equation discretization.} The oracle and the diffusion operator of the circuit are applied $R$ times. Finally, the working qubits are measured to get the desired record from the database. 

Note that multi-controlled $\mathbf{X}$ and $\mathbf{Z}$ gates appear in the circuit. Techniques for decomposing such gates into more elementary ones are demonstrated in \cite{misc-elementary-gates}. However, in practice we can use the implementation of the Grover algorithm available in the Qiskit libraries.

\subsubsection*{Grover Algorithm Adaptation for QUBO}

Having introduced the general Grover algorithm, we will now outline its adaptation for solving QUBO problems. Assume we are solving a problem with $n$ variables and we have $m$ qubits for storing the respective objective function values. Those $n+m$ working qubits are initialized to the quantum state
\begin{equation}
|\psi_{\text{init}}\rangle = \frac{1}{\sqrt{2^n}}\sum_{i=0}^{2^n-1}|i\rangle _n |f(i)-y\rangle_m,
\label{alg_grover_qubo_init}
\end{equation}
where $|i\rangle$ is the $i$\textsuperscript{th} basis state of an $n$-qubit system, $f(i)$ is the objective function value for the bit string encoded in state $|i\rangle$ and $y$ is the lowest value of the objective function found so far. Unless such $y$ exists, it is initialized to $y:=f(s)$, where $s$ is a random $n$ bits long string. 

In fact, state \eqref{alg_grover_qubo_init} is a uniformly distributed superposition consisting of all the possible arguments of the objective function and the respective function values. Note that qubits $|f(i)-y\rangle_m$ are entangled with qubits $|i\rangle _n$ to preserve the relation between function arguments $i$ and function values $f(i)$. It is worth noting that the parameters of the quantum gates in the circuit preparing state $|\psi_{\text{init}}\rangle$ are based on the coefficients of the objective function rather than on the actual function values. Otherwise, we would need to list all the function values ex-ante and the algorithm would be reduced to the brute force method without any speed-up.

If, for a certain argument $k$, it holds that $f(k) < y$, then state $|k\rangle_n$ should be marked, because we have found a better solution than $y$. This means that the oracle is quantum gate $\mathbf{O}$ defined as
\begin{equation}
\mathbf{O}|i\rangle_n |f(i)-y\rangle_m  = \text{sgn}[f(i)-y]|i\rangle_n |f(i)-y\rangle_m \,\,\, \forall i \in \{0;1 \dots 2^n-1\}.
\end{equation}
After state $|k\rangle_n$ has been marked, its probability amplitude is amplified with the diffusion operator (the same one as shown in Figure~\ref{fig_grover_circuit}) and the state stored in the first $n$ working qubits is measured to get $k$. The function value $f(k)$ is determined and a new quantum state $|\psi_{\text{init}}\rangle$ is prepared, but with $y := f(k)$. The algorithms run until the value of the objective function decreases. As can be seen, this adaptation of the Grover algorithm follows the hybrid approach, as the quantum computer has to be reprogrammed in each iteration. Technical details on the construction of the circuit preparing state $|\psi_{\text{init}}\rangle$ and the design of the oracle $\mathbf{O}$ are presented in \cite{grover-qubo}.  

\clearpage

\section{Optimization of the FX Reserves Currency Composition}
\label{secPractical}
In this part, we will discuss the actual application of quantum computers in portfolio optimization. In particular, we will show the capabilities of the approaches presented in Section~\ref{secAlg} as regards finding the optimal currency composition of the CNB's FX reserves. First, we will provide a detailed description of the optimization problem. After that, we will show how the underlying data for the optimization are derived, and finally we will discuss the results from the technical and briefly from the economic perspectives.

\subsection{Formulation of the Optimization Problem}
\label{subsecPracDetailed}
The goal of the optimization problem is to find the optimal currency composition of the CNB's FX reserves after three major crises of the 21\textsuperscript{th} century: the Great Recession of 2007--2009, the European debt crisis of 2011 and the Covid crisis of 2020. Since three periods are involved, the optimization is dynamic. The optimal compositions should generate the maximum possible return in CZK and at the same time the return volatility and the transaction costs incurred in connection with reserve currency rebalancing should be minimized. We require the sum of the weights of all the currencies involved to equal 1 for each time period. The weights have to be non-negative (i.e.~short positions are forbidden) and less than or equal to one. Apart from these requirements, we impose no further constraint on the currency composition.  We include all the currencies we currently have in the reserves in the optimization regardless of they were part of the reserves in the periods mentioned above. This means that the following currencies are included: AUD, CAD, CNY, EUR, GBP, JPY, SEK, USD and gold.  

Note that for the purposes of this paper, we abstract from the particular asset classes included in the reserves and optimize the currency composition only from the perspective of the FX returns in CZK. We could include individual asset classes (such as government bonds, agency bonds and equities) for each currency in the model, but at the cost of additional variables. This would make the model too complex for solving on a quantum computer. Another possible approach would be two-level optimization, i.e.~first find the optimal currency composition and then find the asset composition for each currency. However, we leave this exercise for further research.

\begin{figure}[h]
\caption{MSCI World Developed Markets Index 2000--2021}
\label{fig_crises}
\begin{center}
\includegraphics[scale = 0.65]{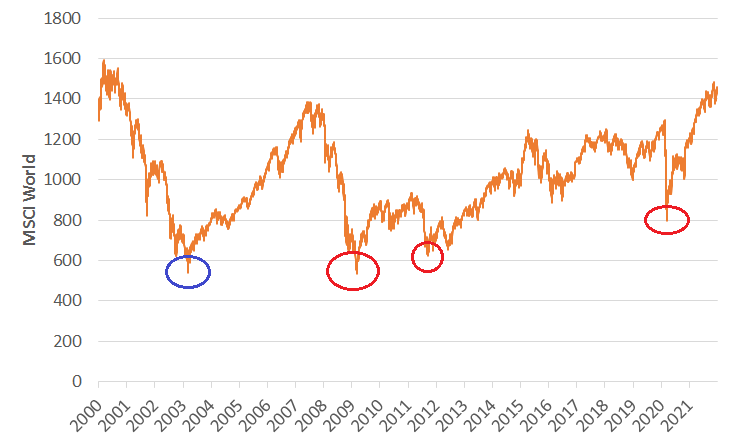}
\end{center}
\vspace{0.1cm}
{\footnotesize \textbf{\textit{Note:}} The troughs of the Great Recession, the European debt crisis and the Covid crisis are highlighted with red circles. Note that we do not include the dot-com crisis of 2002 (the blue circle) in our analysis.}\\
{\footnotesize \textbf{\textit{Source:}} Author's own creation based on Bloomberg data}
\end{figure}

As discussed above, we carry out the optimization for three time periods. To anchor the periods, we looked for troughs (i.e.~local minima) of the MSCI World Developed Markets index. The index comprises more than 1,000 shares listed at exchanges in developed countries.\footnote{Australia, Austria, Belgium, Canada, Denmark, Finland, France, Germany, Hong Kong, Ireland, Israel, Italy, Japan, the Netherlands, New Zealand, Norway, Portugal, Singapore, Spain, Sweden, Swiss, the United Kingdom and the USA.} Thanks to its world-wide coverage, the index is suitable for identifying global crises. The path of the index from 2000 to 2021 is shown in Figure~\ref{fig_crises}. A detailed analysis revealed that the lowest points for the Great Recession, the European debt crisis and the Covid crisis were reached on 9~March 2009, 22~September 2011 and 18~March 2020 respectively. Note that we did not include the year 2022 in our analysis, because the crisis connected with the Russian--Ukrainian conflict is still ongoing.

The above-described version of the optimization problem is the \textit{practical} one. Because of the different capabilities of the individual quantum algorithms and computers we used, we also formulated two simpler versions of the practical problem. The first is a \textit{testing} version comprising only the time period connected with the Great Recession and all nine currencies. The second is a \textit{toy} version comprising the time period connected with the European debt crisis and only three currencies (AUD, CAD and gold). To reduce the number of weights in the toy version further, we removed the constraint $\sum_i w_i = 1$ from the objective function and substituted $w_\text{gold} = 1 - w_\text{AUD} - w_\text{CAD}$  into the function.\footnote{The objective function for the three assets is provided by \cite{misc-markowitz}, page 83.}

Since the optimization is carried out on a quantum computer, the objective function is formulated in QUBO form \eqref{eq_qubo_markowitz_dynamic}. This means that we binarized the asset weights. We used 14, 10 and 3 binary variables per weight for the practical, testing and toy versions respectively. The accuracy of the weights is therefore 0.006\%, 0.1\% and 12.5\% respectively. The total number of binary variables involved is 378, 90 and 6 respectively. The numbers of binary variables used for binarizing the weights come from the maximum number of variables the given algorithm can work with (the toy and testing versions) and from the requirement to ensure sufficient accuracy (the practical version). We set risk aversion $\lambda = 10$, the sensitivity to transaction costs $\mu = 20$ and the importance of the constraint that the sum of the weights is equal to one $F = 100$ (for D-Wave constrained hybrid heuristic $F=0$). These numbers are chosen arbitrarily.

\subsection{Data and Methodology}
\label{subsecPracDataMeth}
The underlying raw data we used are the FX rates between CZK and all the other currencies involved. The rates were expressed in the notation \textit{ccy}CZK, i.e.~the number of CZK units that one unit of the currency \textit{ccy} is worth. The data come from the CNB's FX rates list and Bloomberg. Based on the bid and ask rates we calculated the mid rate.

The daily CZK return on the $i$\textsuperscript{th} currency is given by the expression $100(FX_t^{(i)}/FX_{t-1}^{(i)} - 1)$, where $FX_t^{(i)}$ is the mid rate for day $t$. The daily returns were averaged over a time period starting on the day the trough of the crisis was reached and spanning five years to get an estimation of the expected returns. Note that for the Covid crisis the series ends on 31~December and is therefore only 1.75 years long. The averaged returns were annualized. Based on the daily returns, we calculated the variances in the returns and the covariances among the returns of all currencies. The variances and covariances were also annualized. Transaction costs are based on bid/ask spreads. On each day, for the $i$\textsuperscript{th} currency we calculated the relative spread $100(FX_\text{ASK}^{(i)} - FX_\text{BID}^{(i)})/FX_\text{MID}^{(i)}$. The relative spreads were averaged over the same time spans as the returns. The expected returns, average transaction costs and covariance matrices for all three time periods are provided in Appendix~\ref{appendixRetCostCov}.

\begin{table}[h]
\catcode`\-=12 
\caption{Results of Optimization with MS~Excel Gradient Solver}
\begin{center}
{\footnotesize
\begin{tabular}{l|l|ccccccccc}
\hline\hline
\rule{0pt}{2.5ex} \textbf{Version} & \textbf{Time period} & \textbf{USD} & \textbf{EUR} & \textbf{AUD} & \textbf{CAD} & \textbf{GBP} & \textbf{SEK} & \textbf{JPY} & \textbf{CNY} & \textbf{Gold}  \\ \hline

\rule{0pt}{2.5ex} \multirow{3}{*}{Practical}  & Great Recession & 0.0\% & 31.7\% & 19.8\% & 1.2\% & 5.9\% & 35.6\% & 0.0\% & 0.0\% & 5.7\%\\
\rule{0pt}{2.5ex} & Debt Crisis & 6.2\% & 42.6\% & 15.2\% & 0.0\% & 6.4\% & 23.5\% & 0.0\% & 4.8\% & 1.4\%\\
\rule{0pt}{2.5ex} & Covid & 0.0\% & 20.1\% & 32.7\% & 0.6\% & 12.7\% & 21.2\% & 0.0\% & 3.7\% & 8.9\%\\\hline
\rule{0pt}{2.5ex} Testing & Great Recession & 0.0\% & 26.2\% & 17.1\% & 0.0\% & 0.0\% & 52.9\% & 0.0\% & 0.0\% & 3.8\%\\ \hline
\rule{0pt}{2.5ex} Toy & Debt Crisis &  &  & 31.1\% & 54.7\% &  &  &  &  & 14.2\% \\ 
\rule{0pt}{2.5ex} Toy -- bin & Debt Crisis &  &  & 37.5\% & 50.0\% &  &  &  &  & 12.5\% \\ \hline
	
\end{tabular}}
\end{center}
\vspace{0.1cm}
{\footnotesize \textbf{\textit{Note:}} In the toy version, only AUD, CAD and gold are included. Since only three binary variables are used for the representation of each weight, the optimal solution that can be found in this setting is different from the one obtained using the gradient method. This solution is depicted in row \textit{Toy -- bin}.}\\
{\footnotesize \textbf{\textit{Source:}} Author's own calculations}
\label{tabResStandard}
\end{table}

Based on the expected returns, the average transaction costs and the covariance matrices, we carried out the optimization using a continuous gradient solver implemented in MS~Excel. The results for the practical, testing and toy versions of the optimization are provided in Table~\ref{tabResStandard}. The run time of the gradient method was 2,109, 375 and 94 milliseconds for the practical, testing and toy model respectively.\footnote{Note that the tests of all the classical algorithms were carried out on a PC equipped with 8~GB RAM and a four-core processor with clock speed 3.3 GHz.} The optimal value of the objective function was 0.0921, 0.0177 and 0.088 for the practical, testing and toy model respectively. We find that in all instances, the eigenvalues of the matrix defining the objective function are positive, i.e.~the matrix is positive definite. This means that we are dealing with the ``easy version'' of the quadratic optimization problem, which makes it harder for quantum computers to outperform classical ones. Based on these results, we will assess the capabilities of quantum algorithms in practical applications.

It is important to note that these benchmark values concern portfolio optimization with real asset weights, whereas the other algorithms tested are intended for the QUBO version of the problem. As the purpose of this article is to assess the practical capabilities of quantum algorithms, we compared them with the option most available to bank analysts, the MS~Excel solver, despite the different nature (i.e.~continuous vs binary) of the algorithms.

\subsection{Results}
\label{subsecPracRes}

In this part, we will discuss the results we obtained. First, we will present the results for all three versions of the problem from the perspective of the ability of the algorithms employed to find the optimal solution and the run time. Second, we will briefly discuss the economic interpretation of the optimal results.

Note that in the case of the toy version, we ran the optimization only once, as we immediately obtained the optimal result. In the case of the testing and practical versions, we repeated the optimization 100 times for the D-Wave hybrid constrained heuristic and quantum annealer (in particular, we used the \textit{D-Wave Advantage System 4.1} quantum processor) and 10 times for the classical heuristics, as the run times were much longer in the latter case. After that, we picked the solution nearest to the optimal one in terms of the objective function value. For the branch and bound algorithm, we ran the calculation only once, as the algorithm is exact. Note that for the testing and practical versions, we show the currencies' weights in condensed graphical form; the numerical results are presented in Appendix~\ref{appendixRes}.

\subsubsection*{Toy Version}

The toy version was devised to demonstrate portfolio optimization on IBM Quantum\textsuperscript{TM}, due to the limited number of qubits on the available quantum processors. In particular, we used the Nairobi and Oslo processors, both of which have seven qubits. Such a low number of qubits allowed us to run the portfolio optimization with at most three currencies and to use only three binary variables to represent each currency weight. This would imply nine binary variables, but one weight was expressed with the other two, hence we actually needed six qubits to run the VQE and the QAOA. 

Both the VQE and the QAOA were run on the Nairobi and Oslo real quantum processors with success. As both algorithms are iterative, we compared the number of iterations necessary for real processors and the noiseless simulator. The comparison is depicted in Table~\ref{tabResIbm}. 

\begin{table}[h]
\catcode`\-=12 
\caption{Results of Optimization with QAOA and VQE for Toy Version}
\begin{center}
{\footnotesize
\begin{tabular}{l|c|c}
\hline\hline
\rule{0pt}{2.5ex} \textbf{Processor} & \textbf{VQE} & \textbf{QAOA} \\ \hline
\rule{0pt}{2.5ex} Simulator & 26 & 119 \\
\rule{0pt}{2.5ex} Nairobi & 61 & 118 \\
\rule{0pt}{2.5ex} Oslo & 61 & 154 \\ \hline
	
\end{tabular}}
\end{center}
\vspace{0.1cm}
{\footnotesize \textbf{\textit{Note:}} The table presents the number of iterations needed to find the optimal solution of the toy version. The number of repetitions (shots in IBM terminology) in each iteration was set automatically to 4,000.}\\
{\footnotesize \textbf{\textit{Source:}} Author's own calculations}
\label{tabResIbm}
\end{table}

As can be seen from the table, the number of iterations is higher in the case of real processors. The only exception is the QAOA run on the Nairobi processor, where the number of iterations is lower. We found that the VQE necessitated less iterations than the QAOA. We attribute this better performance to the fact that the VQE is a more general algorithm than the QAOA (the VQE assumes no specific structure of the Hamiltonian describing the optimization problem) and thus the quantum circuit is less complicated. The run time of one iteration of both the VQE and the QAOA is in the order of seconds. Moreover, the batch-processing approach used in the IBM Quantum\textsuperscript{TM} environment increases the total run time to hours. Together with the low number of qubits, this fact disqualifies optimization algorithms designed for gate-based quantum computers from real-world application for the time being. 

The situation was even more complicated in the case of Grover adaptive search. On top of the six qubits used for the binary variables, we also needed one ancilla qubit and qubits for storing the objective function values. After experimenting on a simulator, we found that another six qubits are necessary for this purpose. In total, we would need 13 qubits to run Grover adaptive search. Hence, we tested the algorithm on the simulator only. Note that we achieved the desired results.

We also solved the toy version with D-Wave heuristics and classical methods. The run times are compared in Table~\ref{tabResToy}. The table also contains the classification of the algorithm (continuous/QUBO, exact/heuristic and classical/quantum/hybrid) to show the differences between the approaches. The best algorithm is classical simulated annealing, followed by the D-Wave annealer. Interestingly, the slowest approach is the D-Wave hybrid constrained heuristic. However, we should not draw strong conclusions from the results based on the toy version with only six binary variables. In such cases, the majority of the time is consumed by setting up the problem, with the actual solution taking up only a fraction of the run time. In examples with more variables, the relation is rather the opposite. 

\begin{table}[h]
\catcode`\-=12 
\caption{D-Wave and Classical Method Run Times for Toy Version} 
\begin{center}
{\footnotesize
\begin{tabular}{l|l|r}
\hline\hline
\rule{0pt}{2.5ex} \textbf{Method} &\textbf{Type} & \textbf{Run time (ms)} \\ \hline
\rule{0pt}{2.5ex} Simulated annealing & Classical heuristic QUBO & 8.98  \\
\rule{0pt}{2.5ex} D-Wave annealer\textsuperscript{a} & Quantum heuristic QUBO & 26.01 \\
\rule{0pt}{2.5ex} Brute force method & Classical exact QUBO & 69.93 \\ 
\rule{0pt}{2.5ex} Genetic optimization & Classical heuristic QUBO & 80.56 \\ 
\rule{0pt}{2.5ex} Gradient method\textsuperscript{b} & Classical exact continuous & 94.00 \\ 
\rule{0pt}{2.5ex} Branch and bound & Classical exact QUBO & 375.00 \\ 
\rule{0pt}{2.5ex} D-Wave hybrid & Hybrid heuristic QUBO & 2,993.00\textsuperscript{c} \\ \hline
\end{tabular}}
\end{center}
\vspace{0.1cm}
{\footnotesize \textbf{\textit{Note:}} \\
\textsuperscript{a} The number of calculation repetitions (the \textit{number of reads} in D-Wave terminology) was 1,000.\\
\textsuperscript{b} The gradient method serves as the benchmark.\\
\textsuperscript{c} The run time on the quantum annealer was 63.17 ms; the rest of the time was consumed by classical heuristics.}\\
{\footnotesize \textbf{\textit{Source:}} Author's own calculations}
\label{tabResToy}
\end{table}

\subsubsection*{Testing Version}

\begin{figure}[h]
\caption{Best Solutions for Testing Version}
\label{fig_res_1_per}
\begin{center}
\includegraphics[scale = 0.65]{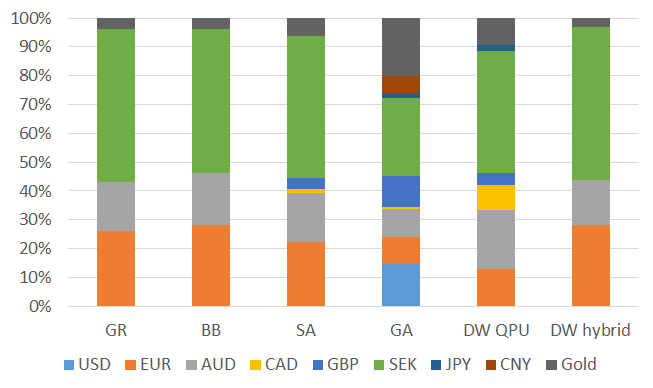}
\end{center}
\vspace{0.1cm}
{\footnotesize \textbf{\textit{Note:}}  \textit{GR} stands for classical gradient algorithm, \textit{BB} for branch and bound algorithm, \textit{SA} for simulated annealing, \textit{GA} for genetic-based algorithm,  \textit{DW QPU} for D-Wave quantum annealer and \textit{DW hybrid} for D-Wave constrained hybrid heuristic. In the case of the D-Wave annealer, the number of repetitions (\textit{reads} in D-Wave terminology) was set to 5,000.}\\
{\footnotesize \textbf{\textit{Source:}} Author's own calculations}
\end{figure}

The testing version comprises 90 binary variables and thus cannot be optimized on IBM Quantum\textsuperscript{TM}. The results obtained with the D-Wave and classical algorithms are compared in Figure~\ref{fig_res_1_per}.

As can be seen in Figure~\ref{fig_res_1_per}, no algorithm was able to find the global optimum. The nearest sub-optimal solution was provided by the D-Wave hybrid heuristic (objective function value 0.01791), followed by the branch and bound algorithm (0.01810), simulated annealing (0.01905), the D-Wave annealer (0.02516) and genetic optimization (0.05578). 

The D-Wave hybrid approach and simulated annealing outperformed the purely quantum D-Wave annealer. We attribute this result mainly to the limited connectivity of the D-Wave quantum processor. The problem has to be adapted to the processor topology, which leads to an increase in the number of ancilla qubits serving as intermediaries in the links between the working qubits. The intermediaries add additional noise and therefore increase the probability of quantum state decoherence. In contrast, the hybrid approach uses the quantum processor for solving less complex sub-problems, which eliminates the need for a large number of ancillas and hence reduces the probability of decoherence. This means that the hybrid approach leverages the capabilities of the quantum annealer as much as possible and at the same time the combination with the classical approach reduces the imperfections of the current quantum hardware.

Despite the fact that the branch and bound algorithm is an exact solver, it was not able to find the optimum. However, the reason is purely technical. The CPLEX\textsuperscript{TM} environment ended with a message that the optimization could not continue and we found that the RAM was full. Increasing the RAM size would allow us to find the global optimum of the problem.

Note that classical genetic optimization was far from the optimal solution. This is probably caused by the fact that, in contrast to simulated annealing, solutions with a higher value of the objective function are always rejected. As a result, the algorithm can get stuck in a local minimum.

The run times of the algorithm were the following: 1,069~ms for the D-Wave annealer, 5,011~ms for the D-Wave hybrid heuristic (of that, 15.232~ms was spent on the quantum annealer), 485 seconds for simulated annealing, 1,382 seconds for genetic optimization and 5,880 seconds for the branch and bound algorithm. This clearly demonstrates the superiority of the D-Wave hybrid approach in the realm of heuristics. In the case of the branch and bound algorithm, despite the very long run time, we would be able to find the global optimum if we had enough memory.

\subsubsection*{Practical Version}

\begin{figure}[h]
\caption{Best Solutions for Practical Version}
\label{fig_res_3_per}
\begin{center}
\includegraphics[scale = 0.65]{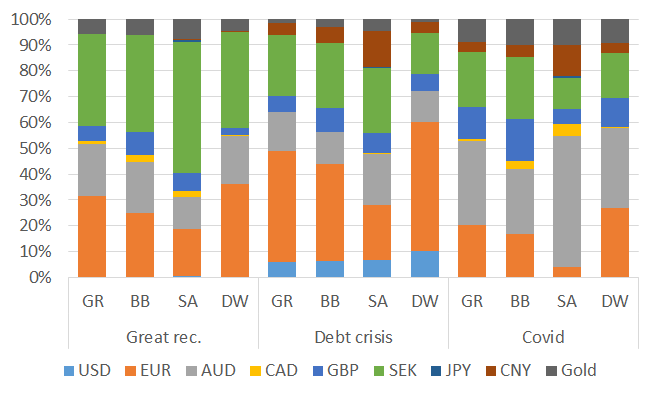}
\end{center}
\vspace{0.1cm}
{\footnotesize \textbf{\textit{Note:}}  \textit{GR} stands for classical gradient algorithm, \textit{BB} for branch and bound algorithm, \textit{SA} for simulated annealing and \textit{DW} for D-Wave constrained hybrid heuristic.}\\
{\footnotesize \textbf{\textit{Source:}} Author's own calculations}
\end{figure}

As the practical version contains 378 binary variables, the D-Wave software was not able to embed the problem in the quantum processor in reasonable time. We also realized that the genetic algorithm is not able to find any feasible solution (i.e.~having a sum of weights equal to 1). Therefore, we only employed simulated annealing, the branch and bound algorithm and the D-Wave hybrid constrained heuristic for the practical version. The solutions obtained are shown in Figure~\ref{fig_res_3_per}.

Similarly to the testing version, no algorithm was able to find the optimal solution. The branch and bound algorithm, the D-Wave hybrid heuristic and simulated annealing found sub-optimal solutions with objective function values of 0.09325, 0.09395 and 0.10710 respectively in 39,429, 5.048 and 6,333 seconds respectively. These results make the D-Wave hybrid heuristic the winner in terms of run time (only 5 seconds). Concerning the closeness to the optimal solution, the branch and bound algorithm achieved the best result, despite its very long run time (almost 11 hours).

The reason why none of the QUBO heuristics tested was able to find the global optimum lies in their nature -- they search in a small subset of all the possible solutions. On the one hand, this allows us to find at least some solution, but on the other hand, we cannot be sure if the solution is optimal. Moreover, binarization of the real weights leads to an objective function with considerably different coefficients, as terms containing $2^{-k}$ have significantly different values for different $k$. In the case of quantum methods, terms with low coefficients can be obscured by noise. In the case of classical heuristics, the algorithm can get stuck in a local minimum despite the measures taken, as in the case of simulated annealing.

Concerning the exact solvers, the branch and bound algorithm again faced the issue of a lack of memory. If we had large enough RAM, we would be able to find the exact solution, but at the cost of a long run time.

\subsubsection*{Economic Interpretation of the Optimal Results}

The actual currency composition of the CNB FX reserves differs significantly from the optimal one that we found for all three time periods. The reserves are composed of all nine mentioned currencies, but the presented optimal solutions contain only some of them.

With the exception of the ``Debt crisis'' time period, the weight of USD is zero. This is caused by its relatively high volatility and low returns in comparison with correlated currencies such as AUD and CAD. However, the American dollar is the primary reserve currency and the prominent currency in international trade. Moreover, the markets for AUD and CAD government bonds are much smaller and less liquid than the US one. Therefore, USD has its place in the FX reserves despite the mentioned disadvantages.

Because of the lower volatility and higher returns of CNY and the high correlation between CNY and JPY, the optimization algorithms prefer the Chinese yuan to the Japanese yen (the weight of the yen is zero). However, until recently, investments in CNY were constrained by the Chinese authorities. For this reason, JPY has been more attractive as a reserve currency.

The Swedish krona and the euro are the currencies with the lowest correlation with the others, so the optimization algorithms increase their weight significantly in order to diversify the reserves. A high share of EUR makes sense, as the Czech Republic is part of the EU and the EUR market is deep and highly liquid. In contrast, the Swedish market is the smallest of those included in our analysis. The share of SEK should be therefore reduced and the share of EUR even increased.

These three examples clearly illustrate the need to provide optimization algorithms with additional information, for example maximum (or conversely  minimum) shares for currencies that suffer from some deficiencies (or conversely are too important to be omitted). Such information has to be based on expert judgment.\footnote{As \cite{misc-markowitz} puts it: \textit{To use the E-V [expected return-variance] rule in the selection of securities we must have procedures for finding reasonable $\mu$ and $\sigma$. These procedures, I believe, should combine statistical techniques and the judgment of practical men. My feeling is that the statistical computations should be used to arrive at a tentative set of $\mu$ and $\sigma$. Judgment should then be used in increasing or decreasing some of these $\mu$ and $\sigma$ on the basis of factors or nuances not taken into account by the formal computations}.} However, we leave more practical optimization for further research once quantum computers are mature enough to easily incorporate constraints into optimization problem.

Another issue is connected with the estimation of the covariance matrices, expected returns and average transaction costs. To get a meaningful estimation, relatively long time series are needed. For example, we tried to estimate the covariance matrix for the Great Recession period with a one-year-long series. However, the optimization ended with 100\% allocation to the Australian dollar. Clearly, such portfolio is poorly diversified, and we had to prolong the time series to five years. This led to another problem, as the five-year time series covers not only the recovery after the Great Recession, but also part of the European debt crisis. On top of that, we have to bear in mind that our analysis is retrospective. In practice, we would need to carry out a forward-looking estimation of the covariances and returns, because we would be interested in the future currency composition of the reserves. In the end, we put these deficiencies of our model aside, because the main purpose of this article is to examine the capabilities of quantum optimization algorithms. We leave a discussion of the estimations of inputs to portfolio optimization to further research.

\section{Conclusion}
\label{secConcl}
The main aim of this paper was to assess the capabilities of quantum algorithms for portfolio optimization and compare them with their classical counterparts. In particular, we tested algorithms for universal gate-based quantum computers (the VQE, the QAOA and Grover adaptive search) on IBM Quantum\textsuperscript{TM}, the quantum annealers and hybrid heuristics provided by D-Wave, the classical branch and bound algorithm (implemented in IBM CPLEX\textsuperscript{TM}) and two classical heuristics (simulated annealing and a genetic-based algorithm). We tested the algorithms from the perspective of run time and ability to find the optimal solution. To carry out the analysis, we constructed a binary version of Markowitz dynamic portfolio optimization in order to find the optimal currency composition of the CNB's FX reserves. The continuous gradient method implemented in MS~Excel served as a benchmark for the purposes of comparison, as the original version of Markowitz optimization is continuous.

We found that none of the QUBO algorithms tested was able to find the optimal solution discovered by the classical continuous gradient method implemented in MS~Excel, although the D-Wave hybrid constrained heuristic was very close in the case of the testing version of the problem, as was the branch and bound algorithm in the case of the practical version. However, this all means that the QUBO approach seems not to be a suitable option for portfolio optimization. Even with a usual office PC, we are able to find the global optimum of the problem in seconds using the classical continuous gradient method implemented in MS~Excel. For the purposes of central banks, this is plainly sufficient. The QUBO version of portfolio optimization is useful only when the problem is purely binary, i.e.~when we only want to decide whether or not to incorporate certain assets into the portfolio and we do not need to know the weights of those assets.

However, this conclusion does not mean that quantum QUBO algorithms are useless. As mentioned earlier, they can be used to solve problems which are binary by definition (such as the travelling salesperson problem or the assignment problem), but we leave this to further research. Moreover, the inability of the algorithms to find the solution to the problem could mean that the problem is hard to solve (we can see this from the excessive run time of the exact branch and bound algorithm). Therefore, it can be used as an interesting testing problem for new algorithms or for measuring progress in the development of quantum computers.

Concerning the capabilities of quantum annealers, we found that the current noisy quantum computers are not able to solve problems of practical size themselves, but they can be used to enhance classical algorithms. The D-Wave hybrid heuristic seems to be just one step from real-world application. Although this algorithm was not always the fastest, it was able to find the solution closest to the optimal one in reasonable time, especially when compared with classical algorithms, both heuristic and exact.

Regarding the QAOA, the VQE and Grover adaptive search, we will have to wait until the number of qubits offered by universal gate-based quantum computers is high enough before we can assess their ability to solve problems of practical size. Grover adaptive search has an even longer road ahead, as the consumption of qubits is higher than in case of the QAOA and the VQE. Moreover, as the Grover algorithm has guaranteed speed-up, we suspect that qubit quality will also be an important factor in its performance.

For the time being, this article should serve as a complement to the introductory literature on quantum computing for economists in general and central bankers and regulators in particular. We expect that it will take several years for quantum computers to become a fully mature technology. Moreover, it is possible that Lloyd's conjecture on exponential speed-up for quantum system simulation will be proved. This would mean that quantum computers will be helpful in solving QUBO problems. In the meantime, we will ``keep an eye'' on the development of quantum computers and certainly return to this research once they are on the verge of industry-scale deployment.

\vspace{0.2cm}
\noindent
\fbox{%
\parbox{\textwidth}{
\textbf{Disclaimer}
\vspace{0.1cm}

This article is intended to be a purely academic research and educational material. The article is neither an offer to buy or sell any security nor a solicitation to do so. The article is not investment advice. Neither the author nor the CNB is responsible for any losses (including but not limited to financial and reputational ones) incurred in connection with the implementation of the methods described in the article. The author declares that no result presented in the article has been implemented as a part of the CNB's investment strategy or used as an input to its strategic asset allocation decision-making. The use of the software products discussed in the article is non-profit-oriented (research and educational purposes).
}}

\clearpage
\nocite{*}
\bibliographystyle{ieeetr}
\bibliography{ms}

\clearpage
\appendix

\section{Matrix Formulation of the Portfolio Optimization Problem}
\setcounter{table}{0}
\setcounter{figure}{0}
\setcounter{equation}{0}
\renewcommand{\thetable}{A\arabic{table}}
\renewcommand{\thefigure}{A\arabic{figure}}
\renewcommand{\theequation}{A\arabic{equation}}

\label{appendixMatrixObj}

In this appendix, we present the matrix form of the objective function \eqref{eq_qubo_markowitz_dynamic}. We will use symbol $\mathbf{1}$ for a matrix or vector (depending on the context) in which all the elements are equal to 1. Symbol  $\mathbf{O}$ is used for a zero matrix. We assume risk aversion $\lambda$ and sensitivity to transaction cost $\mu$ to be time-independent. Note that in contrast to part~\ref{secQuboPtf} we use Dirac notation for linear algebra objects to make the expressions easily readable.

First, we rewrite the objective function in terms of weights $w_i \in \mathbb{R}$ and after that we binarize the function with binary variables $x_i \in \{0;1\}$ to get the QUBO form of the problem.

Condition $\sum_i w_i^{(t)} = 1 \,\,\, \forall t \in \{1,2,\dots T\}$ is expressed as 

\begin{equation}
F\big(\sum_i w_i^{(t)} - 1\big)^2 = F\Big[\big(\sum_i w_i^{(t)}\big)^2 - 2\sum_i w_i^{(t)}+1\Big], 
\end{equation}
in matrix form  as
\begin{equation}
F(\langle w^{(t)}|\mathbf{1}|w^{(t)}\rangle - 2\langle w^{(t)}|\mathbf{1}\rangle + 1).
\end{equation}

Transaction costs at time $t$ are given by the expression $\sum_i \nu_i^{(t)}(w_i^{(t)} - w_i^{(t-1)})^2$.  Considering $\mathbf{V_t}$ to be a diagonal matrix containing unit transaction cost (the $i$\textsuperscript{th} diagonal element is $\nu_i^{(t)}$), the matrix form of the transaction cost term is 
\begin{equation}
\langle w^{(t)} - w^{(t-1)} | \mathbf{V_t} |w^{(t)} - w^{(t-1)} \rangle = \langle w^{(t)} |\mathbf{V_t} |w^{(t)}\rangle + \langle w^{(t-1)} |\mathbf{V_t} |w^{(t-1)}\rangle - \langle w^{(t-1)} |\mathbf{V_t} |w^{(t)}\rangle - \langle w^{(t)} |\mathbf{V_t} |w^{(t-1)}\rangle. 
\end{equation}
Note that $\mathbf{V_{T+1} = \mathbf{O}}$, because we assume no further changes in the portfolio after time $T$.

Together with covariance matrix $\mathbf{C_t}$, the quadratic part of the objective function referring to time period $t$ becomes $\mathbf{A_t} = \lambda\mathbf{C_t} + \mu(\mathbf{V_t} + \mathbf{V_{t+1}}) + F\mathbf{1}$. Besides this, we have mixed terms $ \mu\langle w^{(t-1)} |\mathbf{V_t} |w^{(t)}\rangle$ and  $\mu\langle w^{(t)} |\mathbf{V_t} |w^{(t-1)}\rangle$ connecting consecutive time periods. Putting all that together, we arrive at the final matrix describing the quadratic part of function \eqref{eq_qubo_markowitz_dynamic} 

\begin{equation}
\mathbf{A} =
\begin{pmatrix}
\mathbf{A_1} & -\mu\mathbf{V_2} & \mathbf{O} & \mathbf{O} & \mathbf{O} & \dots & \mathbf{O} \\
-\mu\mathbf{V_2} & \mathbf{A_2} & -\mu\mathbf{V_3} & \mathbf{O} & \mathbf{O} & \dots & \mathbf{O} \\
\mathbf{O} & -\mu\mathbf{V_3}  &\mathbf{A_3} &  -\mu\mathbf{V_4} & \mathbf{O} & \dots & \mathbf{O} \\
\dots & \dots & \dots & \dots & \dots & \dots & \dots\\
\dots & \dots & \dots & \dots & \dots & \dots & \dots\\
\mathbf{O} & \mathbf{O} & \dots & \mathbf{O} & -\mu\mathbf{V_{T-1}} & \mathbf{A_{T-1}}& -\mu\mathbf{V_{T}} \\
\mathbf{O} & \mathbf{O} & \dots &  \mathbf{O} &  \mathbf{O} & -\mu\mathbf{V_{T}}  & \mathbf{A_T} 
\end{pmatrix}
\end{equation}

The linear part of function \eqref{eq_qubo_markowitz_dynamic} consists of asset returns and the linear term of  $\big(\sum_i w_i^{(t)} - 1\big)^2$, therefore $\langle b_t| = -\langle r_t | - 2F\langle\mathbf{1}|$, hence the whole vector $\langle b|$ is $\begin{pmatrix}\langle b_1| & \langle b_2| & \dots & \langle b_T|\end{pmatrix}$. The constant part $c_t = F$ is generated by the absolute term of $F\big(\sum_i w_i^{(t)} - 1\big)^2$. Since we assume $T$ time periods, in total $c = FT$. All these steps allow us to express function \eqref{eq_qubo_markowitz_dynamic} in the form $\langle w| \mathbf{A} |w\rangle + \langle b|w\rangle + c$. 

The next step involves replacing weights $w_i$ with binary variables $x_i$. It holds that $w_i^{(t)} = \sum_{k=1}^\ell 2^{-k}x_{i}^{(k,t)}$. Introducing vector $\langle s| =\begin{pmatrix}2^{-1} & 2^{-2}& \dots & 2^{-\ell}\end{pmatrix}$, we can write  $x_i^{(t)} = \langle s| x_i^{(t)} \rangle$, where $|x_i^{(t)}\rangle$ is a vector of binary variables describing weight $w_i^{(t)}$. Quadratic terms such as $w_i^{(t)} w_j^{(t)}$ can clearly be rewritten as $\langle x_i^{(t)}|s\rangle \langle s|x_i^{(t)}\rangle$. Expression $|s\rangle \langle s| := \mathbf{Q}$ is a matrix of type $\ell \times \ell$. With these new symbols, we get the final shape of the ``binarized'' objective function. In matrix $\mathbf{A}$, we replace each element $a_{ij}$ with matrix $a_{ij}\mathbf{Q}$ and we substitute $b_i \langle s|$ for each element in vector $\langle b|$. Constant $c$ is unaffected by the binarization.

\clearpage
\section{Simulated Annealing for Binary Optimization}
\setcounter{table}{0}
\setcounter{figure}{0}
\setcounter{equation}{0}
\renewcommand{\thetable}{B\arabic{table}}
\renewcommand{\thefigure}{B\arabic{figure}}
\renewcommand{\theequation}{B\arabic{equation}}

\label{appendixSrcSimAnneal}

In this appendix, we provide details on the simulated annealing implementation we used in the practical part of this paper (the MatLab source code is provided in \cite{classical-sa-mtl-implement}). As noted in the general description of the algorithm in Section~\ref{subsecAlgClassical}, the particular implementation of simulated annealing is characterized by a specific noise operator and a cooling schedule. Our noise operator consists of two steps:
\begin{enumerate}
\item Generate integers $i$ and $j$ from a uniform distribution, and swap the values of $x_i$ and $x_j$.
\item Generate random numbers $p_i, p_j \sim U(0,1)$. If $p_i > 0.5$, then $x_i:= \overline{x_i}$, otherwise $x_i$ remains unchanged. Do the same for $j$.
\end{enumerate}

The cooling schedule is defined as follows. First, the temperature is set to $T = 1$. For concrete temperature $T$, the noise operator is applied several times (the maximum number of iterations is defined by the user). The temperature is decreased by a user-defined decrement once the maximum number of iterations is reached. The process repeats until $T>0$.

Sometimes, a solution with a worse value of the objective function is allowed to avoid getting stuck in a local minimum.\footnote{Note that we use the term local minimum loosely. Since the QUBO problem is a discrete problem, any solution is a local minimum. However, in cases where the objective function value is low for a particular solution in comparison with other ones, the algorithm can become trapped in that solution. This is the actual meaning we assign to \textit{get stuck in a local minimum}.} A decision whether to preserve the worse solution is made based on the \textit{Metropolis criterion} \cite{classical-annealing-continuous}. First, the following value is calculated
\begin{equation}
q = \text{exp}\Big({\frac{f_\text{prev} - f_\text{curr}}{k_B T}}\Big),
\end{equation}
where $f_\text{prev}$ and $f_\text{curr}$ are the previous and current values of the objective function respectively, $T$ is the temperature and $k_B > 0$ is the ``Boltzmann constant''. Second, a random number $p  \sim U(0,1)$ is generated and if $q > p$, the worse solution is preserved. Note that in the default setting $k_B = 1$, but the user can alter the constant.\footnote{The actual value of the Boltzmann constant arising in thermodynamics is $k_B = 1.380649 \times 10^{-23} \text{JK}^{-1}$. However, due to the magnitude of the objective function values involved in portfolio optimization, the natural Boltzmann constant is too small.} Increasing the value of  $k_B$ leads to a higher probability of preserving the worse solution. Note that allowing the worse solution to be preserved requires storing the best solution found so far.

The implementation described above was tested on two types of binary optimization problem:
\begin{enumerate}
\item Linear objective functions $\langle b | x \rangle \rightarrow \min$ with $b_i \sim U(-1,1)$. Clearly, the minimum occurs at the point satisfying $x_i = 1 \,\,\,\forall b_i \le 0$ and $x_i = 0$ otherwise. We tested linear functions with up to 50,000 variables.
\item Quadratic objective functions $\langle x | \mathbf{A} | x \rangle \rightarrow \min$, where matrices $\mathbf{A}$ were taken from the library of testing functions for QUBO provided by \cite{classical-testing-functions}. We tested functions with up to 500 variables. In particular, we used the following problems from the library: \textit{be100.1}, \textit{be200.8.1}, \textit{be250.1}, \textit{gka1f} and \textit{gka5f}.
\end{enumerate}

In Table~\ref{tab_sa_tests}1, we provide execution times for all the tested instances to document the capabilities of our implementation. For all the problems tested, our simulated annealing algorithm found the solution. It is worth emphasizing that the algorithm parameters (i.e.~the number of iterations, the temperature-decreasing speed and the ``Boltzmann constant'') had to be fine-tuned in order to achieve the optimum for a particular objective function. There is no scientific approach to setting the parameters; it is rather a matter of experience and intuition. However, two rules of thumb can be used:
\begin{enumerate}
\item With an increasing number of variables, the maximum number of iterations and the ``Boltzmann constant'' should be increased, while the temperature should decrease more slowly. This gives the algorithm enough time to scan a higher number of bit strings in the objective function domain.
\item With an increasing value of the objective function, the ``Boltzmann constant'' should be increased to properly scale the term $f_\text{prev} - f_\text{curr}$ in the Metropolis criterion.
\end{enumerate}

\begin{table}[H]
\label{tab_sa_tests}
\catcode`\-=12 
\caption{Results of Simulated Annealing Implementation Tests}
\begin{center}
{\footnotesize
\begin{tabular}{r|c|rr|l}
\hline\hline 
\multicolumn{1}{c|}{\multirow{2}{*}{\thead{\bf Number of \\ \bf variables}}} & \multirow{2}{*}{\thead{\bf Optimum \\ \bf found }} &  \multicolumn{2}{c|}{\thead{\bf Execution time}}  & \multicolumn{1}{c}{\multirow{2}{*}{\thead{\bf Additional \\ \bf details}}}
\\  \cline{3-4} 
	&  	&	\thead{\bf SA} & \thead{\bf Brute force (est.)} & \\ \hline
	
\multicolumn{5}{c}{\thead{\it Random linear objective function}} \\ \hline

\rule{0pt}{2.5ex} $50$ & yes & $0.08$ s & $13$ d & \\
\rule{0pt}{2.5ex}$100$ & yes & $0.25$ s & $\approx 10^{13}$ y & \\ 
\rule{0pt}{2.5ex}$1000$ & yes & $1.90$ s & $\approx 10^{284}$ y & \\  \hdashline
\rule{0pt}{2.5ex} \multirow{2}{*}{$10000$} & no & $36$ s & \multirow{2}{*}{$\approx 10^{2993}$ y} & opt: $-2 516.8697$, found: $-2 516.8497$ \\ 
\rule{0pt}{2.5ex}	 & yes & $15$ m $38$ s &  & \\  \hdashline
\rule{0pt}{2.5ex} \multirow{2}{*}{$50000$} & no & $1$ m $23$ s & \multirow{2}{*}{$\approx 10^{15034}$ y} & opt: $-19 856.4264$, found: $-19 854.8727$ \\ 
\rule{0pt}{2.5ex}	 & yes & $3$ h $7$ m &  & \\ 

\hline \multicolumn{5}{c}{\thead{\it Quadratic objective function}} \\ \hline
\rule{0pt}{2.5ex} $100$ & yes & $0.18$ s & $\approx 10^{13}$ y & \\ 
\rule{0pt}{2.5ex}$200$ & yes & $2.00$ s & $\approx 10^{43}$ y & \\ 
\rule{0pt}{2.5ex}$250$ & yes & $4.20$ s & $\approx 10^{58}$ y & \\  \hdashline
\rule{0pt}{2.5ex} \multirow{2}{*}{$500$} & yes & $13$ s & \multirow{2}{*}{$\approx 10^{134}$ y} & sparse matrix (10\% non-zero elements) \\ 
\rule{0pt}{2.5ex}	 & yes & $24$ s &  & full matrix \\  \hline

\end{tabular}}
\end{center}
\vspace{0.1cm}
{\footnotesize \textbf{\textit{Note:}} The tests were carried out on a PC equipped with 8 GB RAM and a four-core processor working at 3.3 GHz. As no optimization for parallel computing was implemented, effectively only one core was used. The estimation of the brute force method execution time was based on the assumption that the calculation of the objective function value for a particular bit string and the comparison of the value obtained with the best one found so far takes one nanosecond ($10^{-9}$ s). The assumption is based on the gigahertz ($10^9$ Hz) speed of the CPU clock. As more than one instruction (or processor clock tick) is needed for the described evaluation, our estimations are probably overly optimistic. However, they well document the need for more sophisticated methods for finding the optimum of exponentially scaling problems. Note that the age of the Universe is around $13.77 \times 10^9$ years.}\\
{\footnotesize \textbf{\textit{Source:}} Author's own calculations}
\end{table}

\clearpage
\section{Expected Returns, Transaction Costs and Covariance Matrices}
\setcounter{table}{0}
\setcounter{figure}{0}
\renewcommand{\thetable}{C\arabic{table}}
\renewcommand{\thefigure}{C\arabic{figure}}

\label{appendixRetCostCov}

\vspace{0.2cm}

\begin{table}[H]
\textbf{Great Recession}
{\scriptsize
\begin{center}
\begin{tabular}{l|ccccccccc}
\hline\hline 
\rule{0pt}{2.5ex}   & \textbf{USD} & \textbf{EUR} & \textbf{AUD} & \textbf{CAD} & \textbf{GBP} & \textbf{SEK} & \textbf{JPY} & \textbf{CNY} & \textbf{Gold}  \\ \hline \hline
\rule{0pt}{2.5ex}{\it Returns} & -0.80\% & 0.05\% & 5.00\% & 1.87\% & 2.02\% & 5.90\% & -1.10\% & 1.73\% & 6.28\%\\
\hline
\rule{0pt}{2.5ex}{\it Costs} & 0.26\% & 0.27\% & 0.23\% & 0.22\% & 0.21\% & 0.27\% & 0.20\% & 0.23\% & 0.25\%\\
\hline
\rule{0pt}{2.5ex}{\it CovVar} &  &  &  &  &  &  &  &  & \\
\rule{0pt}{2.5ex}\textbf{USD}  & 1.90\% & 0.67\% & 0.77\% & 1.11\% & 1.20\% & 0.44\% & 1.81\% & 1.89\% & 1.24\%\\
\textbf{EUR}    & 0.67\% & 0.48\% & 0.36\% & 0.45\% & 0.54\% & 0.35\% & 0.74\% & 0.67\% & 0.53\%\\
\textbf{AUD}   & 0.77\% & 0.36\% & 1.34\% & 0.94\% & 0.73\% & 0.53\% & 0.69\% & 0.76\% & 0.90\%\\
\textbf{CAD}   & 1.11\% & 0.45\% & 0.94\% & 1.28\% & 0.84\% & 0.51\% & 0.96\% & 1.10\% & 0.95\%\\
\textbf{GBP}   & 1.20\% & 0.54\% & 0.73\% & 0.84\% & 1.29\% & 0.46\% & 1.22\% & 1.20\% & 1.01\%\\
\textbf{SEK}    & 0.44\% & 0.35\% & 0.53\% & 0.51\% & 0.46\% & 0.91\% & 0.41\% & 0.44\% & 0.50\%\\
\textbf{JPY}     & 1.81\% & 0.74\% & 0.69\% & 0.96\% & 1.22\% & 0.41\% & 2.78\% & 1.80\% & 1.51\%\\
\textbf{CNY}   & 1.89\% & 0.67\% & 0.76\% & 1.10\% & 1.20\% & 0.44\% & 1.80\% & 1.91\% & 1.25\%\\
\textbf{Gold}  & 1.24\% & 0.53\% & 0.90\% & 0.95\% & 1.01\% & 0.50\% & 1.51\% & 1.25\% & 3.87\%\\ \hline
\end{tabular}
\end{center}
}
\vspace{0.1cm}

\textbf{European Debt Crisis}
{\scriptsize
\begin{center}
\begin{tabular}{l|ccccccccc}
\hline\hline 
\rule{0pt}{2.5ex}   & \textbf{USD} & \textbf{EUR} & \textbf{AUD} & \textbf{CAD} & \textbf{GBP} & \textbf{SEK} & \textbf{JPY} & \textbf{CNY} & \textbf{Gold}  \\ \hline \hline
\rule{0pt}{2.5ex}{\it Returns} & 6.12\% & 1.75\% & 0.84\% & 0.89\% & 2.51\% & 1.38\% & 0.74\% & 5.36\% & 1.40\%\\
\hline
\rule{0pt}{2.5ex}{\it Costs} & 0.22\% & 0.18\% & 0.31\% & 0.20\% & 0.27\% & 0.19\% & 0.13\% & 0.18\% & 0.17\%\\
\hline
\rule{0pt}{2.5ex}{\it CovVar} &  &  &  &  &  &  &  &  & \\
\rule{0pt}{2.5ex}\textbf{USD}  & 1.25\% & 0.35\% & 0.69\% & 0.87\% & 0.75\% & 0.42\% & 1.02\% & 1.23\% & 0.67\%\\
\textbf{EUR}    & 0.35\% & 0.26\% & 0.24\% & 0.28\% & 0.29\% & 0.24\% & 0.35\% & 0.35\% & 0.26\%\\
\textbf{AUD}   & 0.69\% & 0.24\% & 1.28\% & 0.87\% & 0.63\% & 0.42\% & 0.62\% & 0.71\% & 0.70\%\\
\textbf{CAD}   & 0.87\% & 0.28\% & 0.87\% & 1.12\% & 0.68\% & 0.41\% & 0.69\% & 0.87\% & 0.65\%\\
\textbf{GBP}   & 0.75\% & 0.29\% & 0.63\% & 0.68\% & 1.01\% & 0.39\% & 0.59\% & 0.76\% & 0.42\%\\
\textbf{SEK}    & 0.42\% & 0.24\% & 0.42\% & 0.41\% & 0.39\% & 0.69\% & 0.37\% & 0.42\% & 0.35\%\\
\textbf{JPY}     & 1.02\% & 0.35\% & 0.62\% & 0.69\% & 0.59\% & 0.37\% & 1.73\% & 1.01\% & 0.95\%\\
\textbf{CNY}   & 1.23\% & 0.35\% & 0.71\% & 0.87\% & 0.76\% & 0.42\% & 1.01\% & 1.29\% & 0.67\%\\
\textbf{Gold}  & 0.67\% & 0.26\% & 0.70\% & 0.65\% & 0.42\% & 0.35\% & 0.95\% & 0.67\% & 3.00\%\\ \hline
\end{tabular}
\end{center}
}
\vspace{0.1cm}

\textbf{Covid Crisis}
{\scriptsize
\begin{center}
\begin{tabular}{l|ccccccccc}
\hline\hline 
\rule{0pt}{2.5ex}   & \textbf{USD} & \textbf{EUR} & \textbf{AUD} & \textbf{CAD} & \textbf{GBP} & \textbf{SEK} & \textbf{JPY} & \textbf{CNY} & \textbf{Gold}  \\ \hline \hline
\rule{0pt}{2.5ex}{\it Returns} & -5.72\% & -4.50\% & 4.34\% & -0.25\% & 0.00\% & -1.22\% & -9.35\% & -0.64\% & 4.02\%\\
\hline
\rule{0pt}{2.5ex}{\it Costs} & 0.33\% & 0.21\% & 0.41\% & 0.34\% & 0.25\% & 0.44\% & 0.30\% & 0.29\% & 0.34\%\\
\hline
\rule{0pt}{2.5ex}{\it CovVar} &  &  &  &  &  &  &  &  & \\
\rule{0pt}{2.5ex}\textbf{USD}  & 0.96\% & 0.44\% & 0.30\% & 0.54\% & 0.40\% & 0.24\% & 0.72\% & 0.81\% & 0.39\%\\
\textbf{EUR}    & 0.44\% & 0.35\% & 0.18\% & 0.26\% & 0.23\% & 0.22\% & 0.40\% & 0.39\% & 0.30\%\\
\textbf{AUD}   & 0.30\% & 0.18\% & 0.74\% & 0.47\% & 0.40\% & 0.35\% & 0.22\% & 0.31\% & 0.41\%\\
\textbf{CAD}   &0.54\% & 0.26\% & 0.47\% & 0.70\% & 0.39\% & 0.29\% & 0.33\% & 0.49\% & 0.28\%\\
\textbf{GBP}   & 0.40\% & 0.23\% & 0.40\% & 0.39\% & 0.60\% & 0.25\% & 0.36\% & 0.37\% & 0.30\%\\
\textbf{SEK}    & 0.24\% & 0.22\% & 0.35\% & 0.29\% & 0.25\% & 0.52\% & 0.21\% & 0.25\% & 0.30\%\\
\textbf{JPY}     & 0.72\% & 0.40\% & 0.22\% & 0.33\% & 0.36\% & 0.21\% & 0.86\% & 0.63\% & 0.54\%\\
\textbf{CNY}   & 0.81\% & 0.39\% & 0.31\% & 0.49\% & 0.37\% & 0.25\% & 0.63\% & 0.82\% & 0.47\%\\
\textbf{Gold}  & 0.39\% & 0.30\% & 0.41\% & 0.28\% & 0.30\% & 0.30\% & 0.54\% & 0.47\% & 2.45\%\\ \hline

\end{tabular}
\end{center}
}
\vspace{0.1cm}
{\footnotesize \textbf{\textit{Note:}}  \textit{Returns} stands for expected returns, \textit{costs} for average transaction costs and \textit{CovVar} for covariance matrix.}\\
{\footnotesize \textbf{\textit{Source:}} Author's own calculation based on the CNB's FX rates list and Bloomberg data}
\label{tableShareReturnCovariance}
\end{table}

\clearpage
\section{Numerical Results}
\setcounter{table}{0}
\setcounter{figure}{0}
\renewcommand{\thetable}{D\arabic{table}}
\renewcommand{\thefigure}{D\arabic{figure}}

\label{appendixRes}

\begin{table}[H]
\textbf{Testing Version}
{\scriptsize
\begin{center}
\begin{tabular}{l|ccccccccc}
\hline\hline 
\rule{0pt}{2.5ex}  \textbf{Algorithm}  & \textbf{USD} & \textbf{EUR} & \textbf{AUD} & \textbf{CAD} & \textbf{GBP} & \textbf{SEK} & \textbf{JPY} & \textbf{CNY} & \textbf{Gold}  \\ \hline \hline
\rule{0pt}{2.5ex}  \textcolor{red}{\textbf{Grad.}} & \textcolor{red}{0.0\%} & \textcolor{red}{26.2\%} & \textcolor{red}{17.1\%} & \textcolor{red}{0.0\%} & \textcolor{red}{0.0\%} & \textcolor{red}{52.9\%} & \textcolor{red}{0.0\%} & \textcolor{red}{0.0\%} & \textcolor{red}{3.8\%}\\
\rule{0pt}{2.5ex} \textbf{BB} & 0.0\% & 28.0\% & 18.2\% & 0.0\% & 0.0\% & 49.9\% & 0.0\% & 0.0\% & 3.9\%\\
\rule{0pt}{2.5ex}  \textbf{SA} & 0.0\% & 22.2\% & 17.1\% & 1.4\% & 3.9\% & 49.1\% & 0.2\% & 0.0\% & 6.2\%\\
\rule{0pt}{2.5ex}  \textbf{GA} & 14.6\% & 9.6\% & 9.7\% & 0.5\% & 11.0\% & 27.1\% & 1.7\% & 5.8\% & 20.3\%\\
\rule{0pt}{2.5ex}  \textbf{D-Wave QPU} & 0.2\% & 12.7\% & 20.4\% & 8.6\% & 4.4\% & 42.2\% & 2.1\% & 0.3\% & 9.2\%\\
\rule{0pt}{2.5ex}  \textbf{D-Wave hyb.} & 0.1\% & 28.0\% & 15.5\% & 0.2\% & 0.0\% & 53.0\% & 0.0\% & 0.2\% & 2.9\%\\ \hline

\end{tabular}
\end{center}
}
\vspace{0.3cm}

\textbf{Practical Version}

{\scriptsize
\begin{center}
\begin{tabular}{l|l|ccccccccc}
\hline\hline 
\rule{0pt}{2.5ex}   \textbf{Period} & \textbf{Algorithm}  & \textbf{USD} & \textbf{EUR} & \textbf{AUD} & \textbf{CAD} & \textbf{GBP} & \textbf{SEK} & \textbf{JPY} & \textbf{CNY} & \textbf{Gold}  \\ \hline \hline
\rule{0pt}{2.5ex} \multirow{4}{*}{\textbf{Great Rec.}} & \textcolor{red}{\textbf{Grad.}} & \textcolor{red}{0.0\%} & \textcolor{red}{31.7\%} & \textcolor{red}{19.8\%} & \textcolor{red}{1.2\%} & \textcolor{red}{5.9\%} & \textcolor{red}{35.6\%} & \textcolor{red}{0.0\%} & \textcolor{red}{0.0\%} & \textcolor{red}{5.7\%}\\
\rule{0pt}{2.5ex}  & \textbf{BB} & 0.0\% & 25.0\% & 19.7\% & 2.6\% & 9.0\% & 37.5\% & 0.0\% & 0.0\% & 6.2\%\\
\rule{0pt}{2.5ex} & \textbf{SA} & 0.5\% & 18.4\% & 12.4\% & 2.1\% & 7.2\% & 50.5\% & 0.8\% & 0.6\% & 7.5\%\\
\rule{0pt}{2.5ex} & \textbf{D-Wave hyb.} & 0.1\% & 36.2\% & 18.5\% & 0.3\% & 2.7\% & 37.1\% & 0.1\% & 0.2\% & 4.8\%\\ \hline
\rule{0pt}{2.5ex} \multirow{4}{*}{\textbf{Debt crisis}} & \textcolor{red}{\textbf{Grad.}} & \textcolor{red}{6.2\%} & \textcolor{red}{42.6\%} & \textcolor{red}{15.2\%} & \textcolor{red}{0.0\%} & \textcolor{red}{6.4\%} & \textcolor{red}{23.5\%} & \textcolor{red}{0.0\%} & \textcolor{red}{4.8\%} & \textcolor{red}{1.4\%}\\
\rule{0pt}{2.5ex}  & \textbf{BB} & 6.2\% & 37.5\% & 12.5\% & 0.0\% & 9.4\% & 25.0\% & 0.0\% & 6.3\% & 3.1\%\\
\rule{0pt}{2.5ex} & \textbf{SA} & 6.6\% & 21.4\% & 19.6\% & 0.4\% & 7.9\% & 25.0\% & 0.5\% & 14.2\% & 4.4\%\\
\rule{0pt}{2.5ex} & \textbf{D-Wave hyb.} & 10.1\% & 50.0\% & 12.1\% & 0.1\% & 6.5\% & 15.9\% & 0.0\% & 4.1\% & 1.1\%\\ \hline
\rule{0pt}{2.5ex} \multirow{4}{*}{\textbf{Covid}} & \textcolor{red}{\textbf{Grad.}} & \textcolor{red}{0.0\%} & \textcolor{red}{20.1\%} & \textcolor{red}{32.7\%} & \textcolor{red}{0.6\%} & \textcolor{red}{12.7\%} & \textcolor{red}{21.2\%} & \textcolor{red}{0.0\%} & \textcolor{red}{3.7\%} & \textcolor{red}{8.9\%}\\
\rule{0pt}{2.5ex}  & \textbf{BB} & 0.0\% & 16.9\% & 25.0\% & 3.1\% & 16.4\% & 23.9\% & 0.0\% & 4.7\% & 10.0\%\\
\rule{0pt}{2.5ex} & \textbf{SA} & 0.2\% & 3.9\% & 50.6\% & 4.6\% & 5.7\% & 12.4\% & 0.6\% & 12.2\% & 9.9\%\\
\rule{0pt}{2.5ex} & \textbf{D-Wave hyb.} & 0.0\% & 26.7\% & 31.2\% & 0.2\% & 11.3\% & 17.3\% & 0.0\% & 4.2\% & 9.1\%\\ \hline

\end{tabular}
\end{center}
}

\vspace{0.3cm}

{\footnotesize \textbf{\textit{Note:}} \textit{Grad.} stands for classical gradient algorithm, \textit{BB} for branch and bound algorithm, \textit{SA} for simulated annealing, \textit{GA} for genetic-based algorithm, \textit{D-Wave QPU} for D-Wave quantum annealer and \textit{D-Wave hyb.} for D-Wave constrained hybrid heuristic. The gradient solver is highlighted in red, as it is the benchmark.}\\
{\footnotesize \textbf{\textit{Source:}} Author's own calculation}
\label{tableNumericalResults}
\end{table}

\end{document}